\documentclass[12pt]{article}

\usepackage[utf8]{inputenc}
\usepackage[T1]{fontenc}
\usepackage{lmodern}
\usepackage{amsmath,amssymb,amsthm}
\usepackage{graphicx}
\usepackage{booktabs}
\usepackage{pdflscape}
\usepackage{threeparttable}
\usepackage{longtable}
\usepackage{array}
\usepackage{multirow}
\usepackage{setspace}
\usepackage{geometry}
\usepackage{natbib}
\usepackage{caption}
\usepackage{subcaption}
\usepackage{enumitem}
\usepackage{xcolor}
\usepackage{hyperref}
\hypersetup{colorlinks, citecolor=blue, urlcolor=blue, linkcolor=blue}
\usepackage{url}
\usepackage{float}
\usepackage{placeins}
\usepackage{makecell}
\usepackage{comment}
\usepackage{csquotes}
\usepackage{bookmark}
\usepackage{titling}

\usepackage{tikz}
\usetikzlibrary{positioning,arrows.meta,shapes,calc}

\graphicspath{{figures/}{Figures/}}

\newtheorem{assumption}{Assumption}
\newtheorem{proposition}{Proposition}

\newtheorem{lemma}{Lemma}
\newtheorem{corollary}{Corollary}
\newtheorem{theorem}{Theorem}
\theoremstyle{definition}
\newtheorem{example}{Example}

\newcommand{\E}{\mathbb{E}}

\title{Delegation and Oversight: The Design of AI Products\thanks{\protect\hypertarget{Hfootnote.\theHfootnote}{}\protect\linespread{1}\protect\selectfont I am grateful to the Murphy Institute at Tulane University for its hospitality during my visit. I also thank those, whose anonymous comments, helped me improve the paper. All remaining errors are my own.}}

\author{Johann Caro-Burnett\thanks{\protect\hypertarget{Hfootnote.\theHfootnote}{}\protect\linespread{1}\protect\selectfont Graduate School of Humanities and Social Sciences, Hiroshima University and The Murphy Institute, Tulane University. \href{mailto:johanncb@hiroshima-u.ac.jp}{johanncb@hiroshima-u.ac.jp}.}}
\date{\today}

\begin{document}
\maketitle

\begin{abstract}\footnotesize
\noindent AI products combine richer interfaces with higher autonomy. I study how an AI provider designs and prices a product line for users who differ in their readiness to delegate. Each user privately observes her readiness, selects a product, and interacts with an AI that privately observes its reliability. When the readiness density and reliability hazard rate are non-decreasing, I characterize a globally optimal deterministic menu at every readiness. Interface richness and available autonomy increase weakly with readiness. Active users can share a product only at the top, where the product permits full autonomy and every AI type accepts delegation. 

\bigskip


\bigskip

\end{abstract}

\vfill
\pagebreak

\section{Introduction}

Artificial intelligence (AI) is changing how people find information, produce content, and carry out their work. As AI products become more capable to act on users' behalf, they create new opportunities to delegate the execution of tasks. Choosing an AI product therefore shapes both the work a user can assign and the ways she can guide and oversee it. I study how AI providers choose these two product features, available autonomy and interface richness, using a mechanism-design approach.

I use \textit{interface} broadly to include the graphical and programming tools through which users specify tasks, guide AI action, and oversee execution. Some products provide only a basic text interface. Others offer richer interfaces, including structured input forms, persistent context, permission settings, activity logs, performance dashboards, and approval controls. Products also differ in how much autonomy they permit. Some mainly answer questions or generate text. Others can use tools and applications, access and edit files, browse the internet, write code, and continuously perform complex tasks with less step-by-step human direction. Products commonly described as \textit{agentic AI} combine substantial autonomy with rich interfaces.\footnote{Here, agentic is an AI product's ability to carry out multi-step tasks with limited user direction. A product can have configurations with different degrees of autonomy. Following the economic literature, the model's term \textit{AI agent} refers to a player in the delegation stage, regardless of its degree of autonomy.}

Users differ in their capacity to benefit from delegating work to AI. I call this capacity \emph{readiness} and assume that it is privately observed by users. AI agents, in addition, differ in execution reliability, which each AI privately observes. In particular, even the AI provider cannot fully assess it. Thus, both the user and the AI have private information.

These features create a product-design problem along two dimensions: how much autonomous action the provider makes available and how much interface structure the user receives for directing it. The two dimensions are conceptually distinct. Autonomy reduces the need for user involvement, while a richer interface provides more ways to guide and oversee execution. In principle, users with higher readiness to delegate may need less assistance in specifying their tasks. Providers could therefore offer them products with high autonomy but limited interface richness, while offering low-readiness users more tools for guidance and oversight and less autonomy.\footnote{Two examples illustrate how interface richness and available autonomy can vary separately. A hospital claims-review system could provide advanced features such as templates, retrieval from approved sources, logs, and permissions while requiring human approval. In contrast, a provider could allow a customer-service AI to access databases, edit records, and resolve routine cases independently, while giving users only limited tools to specify, monitor, approve, or configure its work.} In a model of quality and warranties, \citet{matthewsmoore1987} show that buyers with higher valuations can receive lower-quality products.

\begin{figure}[!ht]
\caption{Available autonomy and interface richness in AI product design.}
\centering
\includegraphics[width=0.9\textwidth]{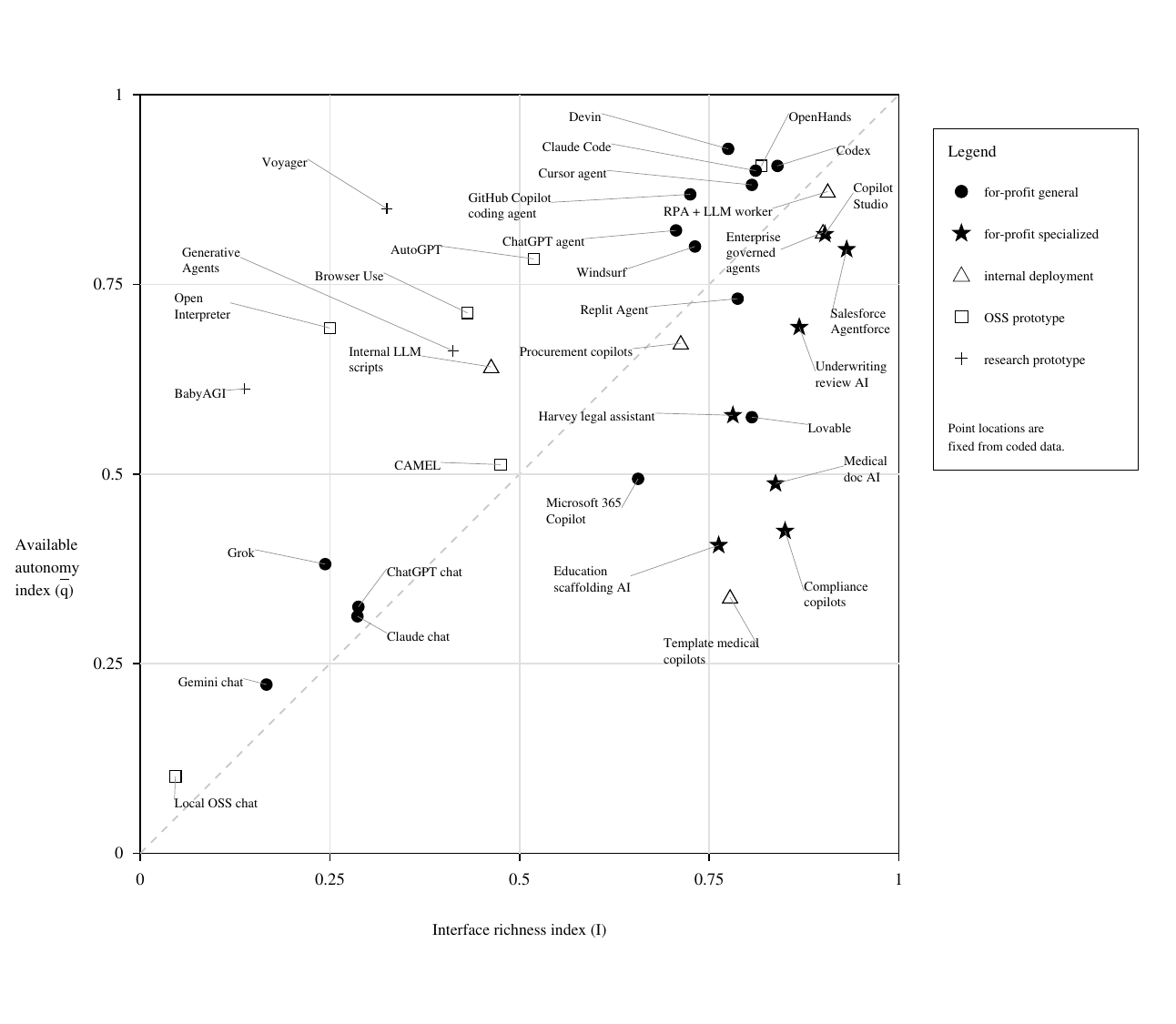}

\begin{minipage}{0.9\textwidth}
\footnotesize
\textit{Notes:} Data from 35 selected AI configurations documented during May 7 and 8, 2026. The AI configurations were sub-classified according to Appendix Table \ref{tab:products}. The horizontal coordinate is an interface richness index and the vertical coordinate is an available-autonomy index. Each index is the average of eight component scores coded with the assistance of ChatGPT. The components are listed in Appendix Tables \ref{tab:Ifeatures} and \ref{tab:Qfeatures}. Appendix Section~\ref{app:coding} describes the coding to obtain the scores in Appendix Table \ref{tab:ai_product_scores}. The correlation between interface richness and available autonomy is 0.55 across all 35 configurations and 0.90 among the 14 for-profit general configurations.
\end{minipage}
\label{fig:ai_product_scatter}
\end{figure}

Nevertheless, widely available commercial AI products tend to bundle higher autonomy with richer interfaces.\footnote{This pattern concerns markets with few providers and many customers. Providers make take-it-or-leave-it offers, leaving users little scope to negotiate individual product features. In bilateral enterprise arrangements, by contrast, large customers may negotiate permissions, integrations, monitoring, autonomy limits, or other features. I model the former market structure below.} Simple products such as chatbots powered by large language models generally combine basic text interfaces with little autonomous execution. In contrast, products with richer tools for interaction and supervision generally permit more autonomous execution.\footnote{For instance, interface features such as persistent context, permission settings, activity logs, and human approval controls often accompany autonomous capabilities such as independently retrieving information, modifying files, executing code, and completing workflows.}
Figure~\ref{fig:ai_product_scatter} illustrates this pattern across providers using selected products and deployments that I coded with AI assistance.\footnote{I selected the examples to illustrate variation in product design. A product is a commercial or open-source AI system, while a deployment is a version configured for a particular organization or workflow. The same product may therefore support deployments with different interfaces, tool access, permissions, approval requirements, monitoring, and autonomy. I use ``deployment'' in Figure~\ref{fig:ai_product_scatter} when the unit is an organizational implementation.} The horizontal axis measures interface richness and the vertical axis measures available autonomy. Off-diagonal points represent products that offer a high level of one feature and a low level of the other.

This pattern is particularly visible among widely used general-purpose AI products sold to individuals and small businesses (the \textit{for-profit general} product label). What leads providers to offer richer interfaces together with higher available autonomy? I study this question within one provider's menu for users in a common task market. I use a mechanism design approach with three participants: an AI provider, a user, and an AI agent. The provider offers products that specify interface richness, an autonomy ceiling, and a price. After selecting a product, the user interacts with an AI that privately observes its task-specific execution reliability. The autonomy ceiling determines what the product permits, while the user and the AI determine how much delegation actually occurs.

The user-AI interaction creates the key delegation trade-off. Depending on the AI's reliability, the user may value delegated execution but may also value retaining control. The AI, however, values execution without accounting for the user's forgone payoff from retaining control. With observable reliability, the user would delegate only to sufficiently reliable AI types. With private reliability and costless reports, the AI either stops or offers to continue. The user therefore decides whether to activate the product's available autonomy for the entire set of AI types willing to act.

The main result characterizes a globally optimal menu at every readiness when the readiness density and the reliability hazard rate are both non-decreasing. Users with low readiness may remain at a common non-delegation baseline. Across active products, interface richness and available autonomy increase weakly with readiness. Active users select distinct products except possibly for one common product at the top. Such pooling can arise only when the product permits full autonomy and every AI reliability type accepts delegation. The characterization covers all deterministic menus under this delegation rule, and the resulting menu satisfies global incentive compatibility.

High-readiness users gain more from products that perform a larger share of their work. As they delegate more, they also gain more from tools that help the AI carry out their intentions accurately. This gives providers a reason to offer higher autonomy together with richer interfaces. However, a richer interface can also make the AI willing to undertake work that the user would still prefer to do herself. Moreover, a user who values delegation highly may still choose a cheaper product if the additional capabilities do not justify its price. The provider therefore chooses interfaces, autonomy, and prices together, taking account of how both users and AI agents respond.

When less reliable AI types begin to accept delegation, their entry can offset the gains from better execution. Thus, the optimal menu may involve pooling or jumps between distinct products.\footnote{For ironing in screening problems, see \citet{myerson1981,toikka2011}. For optimal decision rules with pooling intervals separated by jumps in mechanisms without transfers, see \citet{melumadshibano1991}.} I give conditions on probability distributions and cost curvature that make the active allocation unique and continuous throughout the active region, including where AI acceptance becomes full or autonomy reaches its cap. These conditions allow a jump from the inactive region at the bottom.\footnote{The provider gains from distinguishing users while some AI types still refuse delegation. Once all AI types accept, it can continue to distinguish users through available autonomy. At full autonomy, a richer interface reduces execution losses by the same amount for every user in the pool.} Along a regular branch in the interior of the menu, interface richness, available autonomy, and expected delegation increase strictly with readiness. Higher common AI precision also increases delegation along that branch and can extend service to users with lower readiness. In addition, for sufficiently high common precision, I obtain a global characterization with full AI acceptance under weaker distributional assumptions. Ironing can then generate pooling below the autonomy ceiling.

I build on models in which a firm chooses which products to offer and how to price them \citep{mussa1978,andersoncelik2015}. \citet{yang2025nested} characterizes when a seller offers nested bundles, so each upgrade includes all the goods in the cheaper options. Sellers can distort product characteristics to manage information rents \citep{myerson1981}, and irregular screening problems can generate bunching \citep{rochet1998}.

The model also relates to delegation and communication with privately informed or partially aligned agents \citep{holmstrom1984,dessein2002,alonso2008,deimen2019delegated}. Here, the provider first screens a privately informed user, and the selected product then determines what the AI infers about that user. Interface investment improves execution quality and can also change the set of AI types willing to act. This set affects the user's willingness to pay and the information rents the provider must leave to higher-readiness users. Restrictions on verification and punishments also shape feasible mechanisms \citep{li2020,silva2019}. In this setting, the user screens the AI through the delegation decision, while the provider uses product characteristics to influence both user selection and AI acceptance.

\section{Model}\label{sec:model}

There are three participants: a provider, a user, and an AI. The interaction has two stages. First, the provider offers a menu with prices and the user selects a product. Second, the user and the AI interact under the terms of the selected product to determine delegation.

The delegation stage summarizes a longer interaction in which the AI may carry out a project over several steps. I focus on the decision about the work that remains. After assessing its reliability, the AI indicates only whether it is willing to continue. If it stops, the user retains control of the remaining work. If it is willing to continue, the user chooses how much of that work to delegate, up to the product’s autonomy ceiling.

The user's private readiness type is $x\in[\underline x,\overline x]$, with normalization $1<\underline x<\overline x$. Readiness measures the value of delegation. It may reflect knowledge of the task, the ability to specify objectives and context, the ability to integrate AI work, time savings, and complementarities with later tasks. If the user retains control, she gets a normalized payoff of $1$, while $x$ is the payoff from delegating without errors from the AI. Thus every user prefers error-free delegation to retaining control. User types are distributed according to $G(x)$ with continuously differentiable density $g(x)>0$ on the interior of the support.

The AI's type is $\theta\in[\underline\theta,\overline\theta]$, with $0<\underline\theta<\overline\theta$. I refer to this type as reliability. It measures execution precision for a given task and represents the AI's internal assessment of its expected execution loss. Formally, it is an inverse-variance index. In addition, a common precision parameter $\kappa>0$ scales execution precision for all AI types. Thus $\theta$ captures differences in relative precision across AI agents, while $\kappa$ shifts the overall precision level. AI types are distributed according to $F(\theta)$ with continuously differentiable density $f(\theta)>0$ on the interior of the support. User and AI types are independent, and both distributions are common knowledge.

The provider does not observe private types.\footnote{The provider may know more about the AI than users do and still face uncertainty about its reliability on a particular task.} The user privately observes $x$ before selecting a product, while the AI privately observes $\theta$ and forms a posterior about $x$ from the selected product during the delegation stage. I index the products in a menu $\tau$ by the reported type $\hat x$:
\begin{equation}
\tau(\hat x)
=
\bigl(I(\hat x),\bar q(\hat x),p(\hat x)\bigr),
\end{equation}
where $I>0$ is interface richness, $\bar q\in(0,1]$ is the autonomy ceiling, and $p\ge0$ is the price. The user can instead remain at a costless non-delegation baseline.\footnote{The baseline can include basic tasks such as summarization, simple drafting, or routine retrieval for which reliability is sufficiently predictable. I normalize their value, usage, and common provision cost away. Zero autonomy then means zero incremental delegation of the complex tasks for which reliability remains private.} The selected product $\tau(\hat x)$ is common knowledge during the delegation stage. From a separating product, the AI can infer readiness $\hat x$, which equals $x$ on the truthful path. For a pooled product, the AI instead infers a distribution of readiness.\footnote{Under a deviation to another separating contract, the AI continues to infer the readiness type associated with the selected contract, while the user's payoff depends on her true type. If a range of low-readiness users selects the baseline, the AI's posterior over that range does not affect the allocation because the baseline permits no delegated action. For a pooled product, the AI evaluates acceptance using the expected readiness of the purchasers who would delegate after an acceptance report. If all purchasers delegate, this is their posterior mean readiness. Otherwise, their prospective delegation determines the weights. The AI forms this expectation before observing any action after product selection.}

The provider has additively separable costs
\begin{equation}
K(I)+C(\bar q),
\end{equation}
where both functions are nonnegative, strictly increasing, strictly convex, and twice continuously differentiable on their positive domains. Every $I>0$ is feasible, $K(I)\to\infty$ as $I\to\infty$, and $C$ is continuous on $[0,1]$, with $C(0)=0$.% The interface cost $K$ covers providing and supporting richer interfaces, including context and memory systems, retrieval and structured inputs, monitoring and logs, permissions, evaluations, APIs, administrative tools, and integrations. The autonomy cost $C$ covers making more autonomous action available, including engineering, monitoring, safety, permissions, and liability. I do not model fixed product-development costs separately.

The provider is risk neutral. If a user selects an active product $(I,\bar q,p)$, the provider earns
\begin{equation}
p-K(I)-C(\bar q).
\label{eq:provider_profit}
\end{equation}
%The common baseline yields normalized profit zero. 
The provider chooses the menu to maximize expected profit, anticipating both users' product choices and equilibrium behavior in the subsequent delegation stage.

The sequence of decisions is as follows. The provider first commits to the menu, after which the user observes $x$ and selects a product. The AI observes the selected product% and the task information needed to assess its private $\theta$
, and then forms its posterior about readiness. % I assume that, conditional on the selected product, this task information provides no further information about readiness.
Then, the AI reports whether it is willing to continue. If it stops, the user retains control. If it offers to continue, the user chooses the delegation share $q\in[0,\bar q]$. An execution error is realized on any delegated work, which determines payoffs.

I use perfect Bayesian equilibrium, so the AI updates its beliefs on the equilibrium path using Bayes' rule. The user's choice through a report $\hat x$ therefore determines both the product she receives and what the AI infers about her readiness.
%Whenever an active product supports the positive-delegation continuation described below, I select the continuation with maximal participation: AI types with nonnegative expected payoff accept, and the user follows the stated delegation rule. This selection defines the equilibria I study. Other off-path beliefs can also support no-delegation continuations under perfect Bayesian equilibrium.

I solve the model by backward induction, starting with delegation after the user selects a product $\tau$. Fix a user type $x$ and a selected product $(I,\bar q)$. The user executes her intended action perfectly when she retains control. If she delegates, the AI executes that action with error $\varepsilon$, realized only after the delegation decision. The error has mean zero and a variance that depends on common precision $\kappa$, the AI's type $\theta$, and interface richness $I$:
\begin{equation}
\E[\varepsilon\mid\theta,I]=0
\; \text{ and } \;
\E[\varepsilon^2\mid\theta,I]=\frac{1}{\kappa\theta I}.
\label{eq:error_variance}
\end{equation}
In the hospital claims example, a structured template can reduce errors when the AI extracts billing information, while an approval rule determines which claims it may finalize independently.

Let $q\in[0,\bar q]$ denote the share of the task delegated to the AI, so $1-q$ remains under direct user control. The user's payoff is $1-q+q(x-\varepsilon^2)$, with conditional expectation
\begin{equation}\label{eq:user_payoff}
1-q
+
q\left[x-\frac{1}{\kappa\theta I}\right].
\end{equation}

If the user observed $\theta$, she would delegate up to the available ceiling whenever $\theta$ is at or above the full-information cutoff $\theta_U(x,I):=1/[\kappa I(x-1)]$. Throughout, I extend $F$ below and above its support by $0$ and $1$, respectively, take $f=0$ outside the support, and interpret an integral from a cutoff at or above $\overline\theta$ as an empty tail with value zero. %I use the derivative formulas below only when the relevant cutoff is interior.

The AI privately observes its reliability and sends one of two costless reports. A \textit{no} report ($m=0$) stops delegation, giving $q(x,0)=0$. A \textit{yes} report ($m=1$) is an offer to continue, after which the user chooses $q(x,1)\in[0,\bar q]$. In this stage, there are no transfers or penalties to the AI. The AI's payoff is partially aligned with the user's: both value execution quality, but the AI does not account for the user's forgone payoff from retaining control. Thus, its payoff is $q(x-\varepsilon^2)$. Stopping gives the AI a payoff of zero.\footnote{For instance, an AI refusing delegation may respond, ``I do not have enough reliable information to proceed,'' ``I am not confident I can complete this accurately,'' or ``I can explain how to do this, but I cannot perform the action myself.''} Because the report occurs before $\varepsilon$ is realized, the AI evaluates delegation using its expected payoff.

For the selected product $\tau=(I,\bar q,p)$, let $m_\theta$ denote the equilibrium report of AI type $\theta$, and let $q(x,m)$ denote the user's delegation after report $m$. The AI's incentive-compatibility constraints require, for every type $\theta$ and each report $m\in\{0,1\}$,
\begin{equation}
\mathbb E\left[
q(x,m_\theta)\left(x-\frac{1}{\kappa\theta I}\right)
\middle|\tau
\right]
\ge
\mathbb E\left[
q(x,m)\left(x-\frac{1}{\kappa\theta I}\right)
\middle|\tau
\right].
\label{eq:ai_ic}
\end{equation}
The expectation is over the AI's posterior about readiness after observing the selected product. Because refusal yields zero delegation and zero payoff, these constraints also ensure that participation is optimal for accepting types.

In a separating equilibrium, the selected product reveals $x$. Thus, Equation~\eqref{eq:ai_ic} reduces to
\begin{equation}
\left[q(x,m_\theta)-q(x,m)\right]
\left[x-\frac{1}{\kappa\theta I}\right]
\ge0.
\label{eq:ai_ic_separating}
\end{equation}
Thus, given $(x,I)$, an AI type wants to maximize delegation whenever $x-1/(\kappa\theta I)>0$. Define the resulting acceptance cutoff as $\theta_A(x,I):=1/(\kappa Ix)$. When a \textit{yes} report induces positive delegation, types below $\theta_A(x,I)$ prefer refusal, while types above it prefer to continue. I resolve ties at the cutoff in favor of continuing. If no report induces positive delegation, all AI types are indifferent between reports and I use the same cutoff.

Because $x>1$, $\theta_A(x,I)<\theta_U(x,I)$. In the disagreement region, $\theta\in\bigl(\theta_A(x,I),\theta_U(x,I)\bigr)$, the AI prefers delegation but the user, if reliability were observable, would prefer to retain control. After a \textit{yes} report, the user therefore evaluates the expected payoff across all accepting types. Define the share of AI types willing to accept delegation as
\begin{equation}
\mu(x,I;\kappa)
=
1-F\left(\theta_A(x,I)\right).
\label{eq:mu}
\end{equation}
The user's ex ante marginal gain from increasing delegation to accepting AI types, relative to keeping the task under her control, is
\begin{equation}
D(x,I;\kappa)
=
\int_{\theta_A(x,I)}^{\overline\theta}
\left[x-\frac{1}{\kappa\theta I}-1\right]
f(\theta)\,d\theta.
\label{eq:D}
\end{equation}
When $\mu(x,I;\kappa)>0$, the user's conditional gain per unit of delegation after a \textit{yes} report is $D(x,I;\kappa)/\mu(x,I;\kappa)$. Because her objective is linear in $q$, she chooses $q=\bar q$ if $D(x,I;\kappa)\ge0$ and $q=0$ otherwise, resolving ties in favor of delegation. If $\mu(x,I;\kappa)=0$, the AI refuses with probability one and no delegation occurs.

Under pooling, the AI anticipates that some purchasers may decline to delegate after a \textit{yes} report. Since refusal gives zero, Equation~\eqref{eq:ai_ic} compares $\E[q(x,1)(x-1/(\kappa\theta I))\mid\tau]$ with zero. Whenever $\mathbb E[q(x,1)\mid\tau]>0$, its incentive constraints imply that it accepts if and only if
\begin{equation}
\theta
\ge
\frac{\mathbb E[q(x,1)\mid\tau]}
{\kappa I\,\mathbb E[xq(x,1)\mid\tau]}.
\label{eq:ai_pooled_acceptance}
\end{equation}
From linearity, users still choose either zero or $\bar q$, so the AI weights readiness by prospective delegation in Equation~\eqref{eq:ai_pooled_acceptance}. If no purchaser would delegate, the AI is indifferent between reports.

Thus the AI's acceptance rule and the users' delegation decisions are mutually consistent cutoffs, even in a pooling region. I omit $\kappa$ from function arguments except when defining a function or analyzing its effects.

Consider a user with true readiness $x$ who selects the active separating contract indexed by $\hat x$. The AI observes the contract, infers readiness $\hat x$, and therefore uses the acceptance cutoff $\theta_A\bigl(\hat x,I(\hat x)\bigr)$. The selected contract consequently fixes both the interface and the set of AI types willing to accept delegation.

Conditional on delegation, true type $x$ values execution by each accepting AI type at $x-\hat x$ more than type $\hat x$ does. Therefore, the payoff from selecting contract $\hat x$ is 

\begin{equation}
u(x,\hat x) =1-p(\hat x) + \bar q(\hat x)
\bigl[
D\left(\hat x,I(\hat x)\right)
+\mu\left(\hat x,I(\hat x)\right)(x-\hat x)
\bigr]_+
\label{eq:flat_linear_payoff}
\end{equation}
where $[y]_+=\max\{0,y\}$ and the term $(x-\hat x)$ follows from adding and subtracting $\hat x$ inside the integral in Equation~\eqref{eq:D}. Thus, each active contract gives a payoff that is piecewise affine in true readiness, with slope zero when delegation is off and $\bar q(\hat x)\mu(\hat x,I(\hat x))$ when it is on. 

The provider chooses the menu to maximize expected profit over the full support of readiness. For reports assigned to the baseline, the price and provision cost are zero, and $u(x,\hat x)=1$ for every true type $x$. For separating products, Equation~\eqref{eq:flat_linear_payoff} gives the relevant continuation payoff. For a product that pools a positive mass of readiness types, the equilibrium posterior determines the acceptance set. The provider solves
\begin{equation}
\Pi
= \max_{(I,\bar q,p)}
\int_{\underline x}^{\overline x}
\left[p(x)-K(I(x))-C(\bar q(x))\right]g(x)\,dx,
\label{eq:profit_active}
\end{equation}
subject to global incentive compatibility and participation:
\begin{align}
u(x,x) &\ge u(x,\hat x),
&&\text{for all } x,\hat x\in[\underline x,\overline x],
\label{eq:iclocal}\\
u(x,x) &\ge 1,
&&\text{for all } x\in[\underline x,\overline x].
\label{eq:active_ir}
\end{align}
Equation~\eqref{eq:iclocal} requires each user to prefer the product intended for her type. Equation~\eqref{eq:active_ir} ensures that she receives at least the payoff from the outside option of zero delegation in Equation \eqref{eq:user_payoff}. Along a differentiable active separating menu, Equation~\eqref{eq:iclocal} implies the local first-order condition, but global incentive compatibility requires the additional monotonicity constraint that $\bar q(x)\mu(x,I(x))$ be weakly increasing in $x$.\footnote{Appendix Lemma~\ref{lem:necessary_monotonicity} gives this monotonicity condition. Lemma~\ref{lem:implementation} also considers deviations in which the user turns delegation off. See, for example, \cite{rochet1987rationalizability,ruizdelportal2012conditions}.}

The delegation-stage equilibrium determines the payoff the user anticipates when choosing a product. Let $U(x)=u(x,x)$. Incentive compatibility gives the usual envelope condition: $U(x)$ is absolutely continuous and, for almost every active separating type, $U'(x)=\bar q(x)\mu(x,I(x))$. If the selected product induces no delegation, then $U'(x)=0$. Because delegation and the AI acceptance probability are nonnegative, $U'(x)\ge0$ almost everywhere. To maximize profit, the provider sets prices so that the lowest readiness type receives the baseline payoff, $U(\underline x)=1$. If an inactive interval is present, those users get that same payoff. Therefore,
\begin{equation}
U(x)-1
=
\int_{\underline x}^{x}U'(r)\,dr,
\label{eq:general_rent}
\end{equation}
and expected user information rent satisfies
\begin{equation}
\int_{\underline x}^{\overline x}[U(x)-1]g(x)\,dx
=
\int_{\underline x}^{\overline x}U'(r)[1-G(r)]\,dr.
\label{eq:rent_identity}
\end{equation}

Using Equations~\eqref{eq:profit_active} and \eqref{eq:rent_identity}, I define each type's virtual-surplus contribution in the relaxed separating problem as
\begin{equation}
w(x;I,\bar q;\kappa)
\equiv
\bar q
\left[
D(x,I;\kappa)
-
\frac{1-G(x)}{g(x)}\mu(x,I;\kappa)
\right]
-
K(I)
-
C(\bar q).
\label{eq:w}
\end{equation}
Integrating this contribution against $G$ gives the expected profit for an implemented separating allocation with $U(\underline x)=1$. 

Finally, if readiness $x$ were public, the provider could extract the user's additional expected surplus and would choose $(I,\bar q)$ to maximize $\bar q D(x,I)-K(I)-C(\bar q)$. At an interior solution, $C'(\bar q)=D(x,I)$ and $K'(I)=\bar q D_I(x,I)$. This benchmark removes the first-stage information problem but keeps the second-stage delegation trade-off. The AI still accepts whenever $\theta\ge\theta_A(x,I)$, so AI types in the disagreement region $(\theta_A,\theta_U)$ accept even though the user would retain control if reliability were observable.

\section{The Optimal Product Menu}\label{sec:solution}

The main result allows inactive regions, discontinuities, and pooling. It compares the candidate menu with all deterministic menus under the delegation rules in Section~\ref{sec:model}. I then give conditions for a continuous active menu and study improvements in AI precision $\kappa$.

\begin{assumption}\label{ass:mhr}
The hazard rates $g(x)/[1-G(x)]$ and $f(\theta)/[1-F(\theta)]$ are weakly increasing on the interiors of their supports.
\end{assumption}

Denote $m(x)=(1-G(x))/g(x)$ and $B(x,I)=D(x,I)-m(x)\mu(x,I)$. Thus $m$ is nonincreasing, and $B$ is the virtual return per unit of available autonomy. Define
\begin{equation}
W(x)=\max\left\{0,\sup_{I>0,\,0<\bar q\le1}\left[\bar q B(x,I)-K(I)-C(\bar q)\right]\right\}.
\label{eq:W}
\end{equation}

The zero option is the inactive baseline. For each readiness type, $W(x)$ gives the maximum contribution to profit in the relaxed separating problem. At the endpoints of the readiness support I use the one-sided limits of $m$ and $W$. Under Assumption~\ref{ass:mhr}, the optimal profit $\Pi$ is bounded by the expectation of $W(x)$:
\begin{equation}
\Pi\le\int_{\underline x}^{\overline x}W(x)\,dG(x).
\label{eq:global_profit_bound}
\end{equation}

\begin{lemma}\label{lem:global_bound}
Under Assumption~\ref{ass:mhr}, the set $\{x:W(x)>0\}$ is an upper interval, possibly empty. When nonempty, it has lower boundary $x_0\ge\underline x$, with $x_0=\underline x$ under full coverage. For every readiness type in this interval, at least one product attains the maximum. As readiness increases, the maximizing products have weakly higher interface richness, available autonomy, and effective delegation $\bar q\mu(x,I)$.
\end{lemma}

\begin{proof}
See the \hyperref[proof:global_bound]{Appendix}.
\end{proof}

Pooling changes the AI's beliefs about readiness and therefore its willingness to accept delegation. The proof of Theorem~\ref{thm:global_menu} accounts for these beliefs in deriving Equation~\eqref{eq:global_profit_bound}.

Wherever $W(x)>0$, select the maximizing product (with the highest interface richness, in case of multiplicity) and denote it by $(I^{\ast}(x),\bar q^{\ast}(x))$. Strict convexity of $C$ makes the maximizing autonomy level unique for a given interface, so this rule selects a single product.\footnote{For fixed $\bar q$, the objective has second derivative $\bar q B_{II}(x,I)-K''(I)$. Strict convexity of $K$ alone leaves its sign undetermined.} Choose the baseline wherever $W(x)=0$, including when an active product and the baseline tie at entry.

Expected delegation under this allocation is
\begin{equation}
Q^{\ast}(x)=\begin{cases}
0,&W(x)=0,\\
\bar q^{\ast}(x)\mu(x,I^{\ast}(x)),&W(x)>0.
\end{cases}
\label{eq:Qstar}
\end{equation}

The baseline price is zero. For each $x$ with $W(x)>0$, the price is

\begin{equation}
p^{\ast}(x)=\bar q^{\ast}(x)D(x,I^{\ast}(x))-
\int_{x_0}^{x}Q^{\ast}(t)\,dt.
\label{eq:global_price}
\end{equation}
These prices give the baseline payoff $U(x_0)=1$ at entry and leave higher-readiness users the information rents required by incentive compatibility.

\begin{theorem}\label{thm:global_menu}
Suppose Assumption~\ref{ass:mhr} holds and $p^{\ast}\geq 0$. The resulting menu is globally incentive compatible, individually rational, and globally profit maximizing. Its profit equals the bound in Equation~\eqref{eq:global_profit_bound}.

Up to a set of probability zero, there is at most one active pooling region, an upper interval. On this interval, $\bar q=\mu=1$ and $I$ minimizes
\[
K(I)+\frac{\E[1/\theta]}{\kappa I}.
\]
\end{theorem}

\begin{proof} See the \hyperref[proof:global_menu]{Appendix}.
\end{proof}

As $\mu$ approaches one, less reliable AI types accept delegation, which affects what high-readiness users will pay. The theorem requires nonnegative prices, since a user receiving the baseline payoff could gain by taking a subsidized contract and never activating delegation. Next, I show a sufficient condition for $p^\ast>0$ on the active region.

\begin{corollary}\label{cor:positive_prices}
Suppose Assumption~\ref{ass:mhr} holds and $g$ is non-decreasing. Then $p^\ast > 0$ in the active region. Theorem~\ref{thm:global_menu} therefore applies without an additional price condition.
\end{corollary}

\begin{proof}
See the \hyperref[proof:positive_prices]{Appendix}.
\end{proof}

Under these conditions, the provider can offer the pointwise maximizing products of $W(x)$ and satisfy all user incentives. The menu can contain an inactive lower interval, separating products, and a common top product. While acceptance remains partial, the provider continues to distinguish active users.

Even with this ordering, the menu can jump between distinct separating products. The next assumption gives sufficient conditions for a continuous active allocation.

\begin{assumption}\label{ass:continuity}
The readiness density $g$ is non-decreasing. The reliability density $f$ is twice continuously differentiable on $(\underline\theta,\overline\theta)$, satisfies $\lim_{\theta\downarrow\underline\theta}f(\theta)=0$, and has non-decreasing $\theta^2f'(\theta)$. In addition, the marginal-cost elasticities $\frac{IK''(I)}{K'(I)}$ and $\frac{\bar q C''(\bar q)}{C'(\bar q)}$ are non-decreasing.
\end{assumption}

These conditions strengthen the monotone-hazard assumptions. The vanishing density at the lower bound makes the marginal return to interface richness continuous when AI acceptance becomes full. Together, the density and cost restrictions rule out ties between distinct maximizing products in the active region, so the optimal active menu adjusts continuously as readiness increases.

\begin{theorem}\label{thm:continuous_menu}
Under Assumption~\ref{ass:continuity}, every $x$ with $W(x)>0$ has a unique maximizing product $(I(x),\bar q(x))$. This product varies continuously throughout the active region, including where $\mu(x,I(x))$ reaches one or autonomy reaches its cap. Moreover, $p^\ast(x)>0$ on the active region, and these prices implement a globally optimal allocation that is unique almost everywhere on the readiness support.
\end{theorem}

\begin{proof}
See the \hyperref[proof:continuous_menu]{Appendix}.
\end{proof}

The theorem gives a continuous separating allocation, possibly followed by the terminal pool in Theorem~\ref{thm:global_menu}. Entry from the baseline can involve a jump, and the active allocation can have derivative kinks.\footnote{Appendix Section~\ref{oa:continuity} gives examples of continuous menus.}

\begin{proposition}\label{prop:acceptance_boundary}
Fix readiness $x$ and let ${\underline I}=1/(\kappa x\underline\theta)$. If the lower-end limiting density is finite and positive, then
\begin{equation}
B_I(x,{\underline I}+)-B_I(x,{\underline I}-)
=\frac{[1+m(x)]\underline\theta f(\underline\theta+)}{{\underline I}}>0.
\label{eq:acceptance_jump}
\end{equation}
No positive-autonomy relaxed optimum occurs exactly at ${\underline I}$. Consequently, a continuous active allocation of relaxed maximizers cannot pass from partial to full acceptance when $f(\underline\theta+)>0$.
\end{proposition}

\begin{proof}
See the \hyperref[proof:acceptance_boundary]{Appendix}.
\end{proof}

Proposition~\ref{prop:acceptance_boundary} shows that a positive reliability density at the lower bound creates an upward jump in the marginal return to interface richness when all AI types begin to accept. An active relaxed optimum therefore cannot lie exactly at this boundary, so a menu that moves from partial to full acceptance must jump in interface richness. The condition $f(\underline\theta+)=0$ removes this discontinuity in marginal returns. A positive lower-end density can nevertheless support a continuous menu that stays on one side of the boundary.\footnote{Appendix Section~\ref{app:uniform} gives an explicit uniform-reliability example with a unique smooth, partially accepting, interior product for every readiness type.}

Consider an interval where the selected maximizing product stays on the same smooth branch and has $I^{\ast}>0$, $0<\bar q^{\ast}<1$, an interior AI acceptance cutoff, and a negative-definite Hessian of $w$ with respect to $(I,\bar q)$. I call such an interval regular, with first-order conditions:
\begin{align}
C'(\bar q)&=B(x,I),\label{eq:foc_qbar}\\
K'(I)&=\bar q B_I(x,I).\label{eq:foc_I}
\end{align}
These conditions subtract the information-rent adjustments from the returns to autonomy and interface investment. At the top, $m(x)$ goes to zero, so if this regular interval reaches $\overline x$, the allocation there coincides with the known-readiness benchmark.\footnote{The interface first-order condition requires $D_I(x,I)>0$.} Assumption~\ref{ass:mhr} allows ties between maximizing products. Under Assumption~\ref{ass:continuity}, there is one global maximizer wherever $W(x)>0$, although other products may satisfy the first-order conditions.\footnote{Appendix Section~\ref{oa:selection} shows that, after optimizing autonomy, the profit profile can have a local minimum at a lower interface level that also satisfies the original first-order conditions.}

\begin{proposition}
Under Assumption~\ref{ass:mhr}, on every regular interval,\label{prop:regular_sorting}
\[
I^{*\prime}(x)>0,
\qquad \bar q^{*\prime}(x)>0,
\qquad Q^{*\prime}(x)>0,
\qquad W'(x)>0.
\]
\end{proposition}

\begin{proof}
See the \hyperref[proof:regular_sorting]{Appendix}.
\end{proof}

\begin{corollary}\label{cor:precision_separation}
Under Assumption~\ref{ass:mhr}, for each active type whose selected product stays on the same regular maximizing branch in a neighborhood of $\kappa$, available autonomy, effective precision $\kappa I^{\ast}$, and expected delegation increase strictly with $\kappa$. If the active maximizing branch reaches a locally movable interior entry boundary $x_0$ with zero virtual profit, then
\[
\frac{dx_0}{d\kappa}<0.
\]
\end{corollary}

\begin{proof}
See the \hyperref[proof:precision_separation]{Appendix}.
\end{proof}

Better common AI precision therefore increases delegation along a regular branch and can extend service to users with lower readiness. The increase in effective precision allows $I^{\ast}$ itself to rise or fall. These comparisons follow the same maximizing branch; a switch between distinct maximizers requires comparing their profits.

Sufficiently high common precision allows for a global characterization with full AI acceptance under much weaker distributional assumptions. When all AI types accept and the user delegates, her gain before paying the price is $\bar q[x-1-\E[1/\theta]/(\kappa I)]$. Thus the provider's effective cost combines provision costs with the execution loss that reduces the user's willingness to pay. For $\bar q>0$, define
\begin{equation}
c_\kappa(\bar q)
=C(\bar q)+\min_{I>0}\left\{K(I)+\frac{\bar q\,\E[1/\theta]}{\kappa I}\right\},
\label{eq:high_precision_cost}
\end{equation}
with $c_\kappa(0)=0$. The minimizing interface $I_\kappa(\bar q)$ is unique and satisfies
\begin{equation}
I_\kappa(\bar q)^2K'\!\left(I_\kappa(\bar q)\right)
=\frac{\bar q \E[1/\theta]}{\kappa}.
\label{eq:high_precision_interface}
\end{equation}
Let $\phi(x)=x-1-m(x)$ denote virtual readiness, which subtracts the information-rent adjustment from the value of delegation. Following \citet{myerson1981}, define its ironed version by
\begin{equation}
\bar\phi(x)=\bar\Phi'\!\left(G(x)\right), \text{ where } \Phi(z)=\int_0^z \phi\!\left(G^{-1}(s)\right)\,ds,
\label{eq:ironed_readiness}
\end{equation}
$z\in[0,1]$ is a readiness quantile, and $\bar\Phi$ is the highest convex function that nowhere exceeds $\Phi$ on $[0,1]$. The derivative defines $\bar\phi$ almost everywhere. Convexity makes $\bar\phi$ non-decreasing. On each interval where $\bar\Phi<\Phi$, the slope of $\bar\Phi$ is constant, so $\bar\phi$ is constant over the corresponding readiness interval. These are the ironing intervals.

\begin{proposition}\label{prop:high_precision}
For $\kappa$ above a finite threshold, a globally incentive-compatible and individually rational menu maximizes profit over the stated class of deterministic menus. Its non-decreasing autonomy schedule satisfies
\begin{equation}
\bar q_\kappa(x)\in
\operatorname*{arg\,max}_{0\le\bar q\le1}
\left\{\bar q\bar\phi(x)-c_\kappa(\bar q)\right\}.
\label{eq:high_precision_allocation}
\end{equation}
Choose a common maximizer at each ironed virtual value, so the allocation is constant throughout each ironing interval, and choose the baseline whenever the maximum is zero. Each active type receives interface $I_\kappa(\bar q_\kappa(x))$ and pays
\begin{equation}
p_\kappa(x)
=\bar q_\kappa(x)
\left[x-1-\frac{\E[1/\theta]}{\kappa I_\kappa(\bar q_\kappa(x))}\right]
-\int_{\underline x}^{x}\bar q_\kappa(r)\,dr.
\label{eq:high_precision_price}
\end{equation}
Every AI reliability type accepts every active product in this menu, and every active price is positive. The maximum profit is
\begin{equation}
\Pi
=\int_{\underline x}^{\overline x}
\max_{0\le\bar q\le1}
\left\{\bar q\bar\phi(x)-c_\kappa(\bar q)\right\}\,dG(x).
\label{eq:high_precision_profit}
\end{equation}
Moreover, every optimal equilibrium in this class has full AI acceptance for almost every active user type.
\end{proposition}

\begin{proof}
See the \hyperref[proof:high_precision]{Appendix}. %Equation~\eqref{eq:high_precision_condition} gives a sufficient precision condition.
\end{proof}

Once precision is high enough, the provider can obtain full AI acceptance at a small additional interface cost. Product choice then leaves the AI's acceptance set unchanged, and the provider screens users through their value of delegation. Every active user delegates up to the product's autonomy ceiling, so expected delegation equals $\bar q_\kappa(x)$. Costs and information rents can still exclude users below a readiness cutoff. Full coverage occurs when this cutoff equals $\underline x$.

Equation~\eqref{eq:high_precision_interface} makes interface richness strictly increasing in available autonomy, so both characteristics remain weakly increasing with readiness. The active menu can combine continuous separation, pooling, and jumps. A nonmonotone virtual readiness function can produce pooling through ironing, including at autonomy levels below one. Within each pool, interface richness, autonomy, and price are constant. Because $c_\kappa$ can be nonconvex, the maximizing product can jump as readiness increases, and Equation~\eqref{eq:high_precision_allocation} requires a global maximization. Each active product has an interior interface choice, with $0<I_\kappa(\bar q_\kappa(x))<\infty$, while available autonomy can reach $\bar q=1$. Once autonomy reaches this cap, all higher-readiness users receive the same top product.

\section{Conclusion}

This paper studies how providers design AI products when users privately observe their readiness to delegate and AI agents privately observe their reliability. With a non-decreasing readiness density and a weakly increasing reliability hazard rate, I characterize a globally optimal deterministic menu at every readiness. Interface richness and available autonomy increase weakly with readiness. The provider separates active users except possibly at a common top product with full AI acceptance and full autonomy. Additional restrictions on reliability and cost curvature make the active allocation unique and continuous. When common AI precision is sufficiently high, I also characterize an optimal menu with full AI acceptance even when hazard rates are nonmonotone. In that case, ironing can generate pooling below full autonomy.

Even when every AI type is willing to act, the provider may profit from limiting how much work the AI can perform. Such limits can arise when broader execution is both technically feasible and legally permitted. The cost $C(\bar q)$ can include expected liability and compliance costs alongside engineering, monitoring, and support costs. Offering more autonomy on an active product for lower-readiness users also makes it more attractive to higher-readiness users. To keep them from switching, the provider must leave them higher information rents. These rents give the provider a reason to limit autonomy even when expected liability costs are low. A similar screening motive can make it profitable to reduce the quality of a lower-priced version \citep{deneckeremcafee1996}.\footnote{\citet{crawfordshum2007} estimate quality degradation in cable television packages.}

Beyond setting the autonomy ceiling, a provider could influence when the AI offers to continue by changing how it evaluates the AI's work. In the model, the AI values execution quality without accounting for the user's forgone payoff from retaining control. For example, a hospital could give a claims-review AI credit for referring a claim when a human reviewer would handle it better. A separating product reveals the user's readiness to the AI. If the AI fully accounted for her payoff from retaining control, it would apply the delegation cutoff she would choose if she observed its reliability. At the same interface, autonomy ceiling, and price, the AI could then do less work while raising the user's expected payoff.

Evaluations, monitoring, and prior performance may also give the provider better information about AI reliability than users have. Disclosing that information could change which tasks users delegate and what they are willing to pay for each product. When can a provider credibly communicate a favorable assessment of its AI, and how would such disclosure affect the menu it offers?

Specialized AI products may involve different contracting arrangements. Large enterprise customers can negotiate product characteristics and prices directly with providers. Under complete information, bilateral bargaining can implement allocations that maximize joint surplus, with prices determining its division. A provider could use a common interface architecture across specialized deployments while negotiating autonomy limits that reflect industry-specific requirements, task reliability, and costs. This could explain why specialized products with similarly rich interfaces permit different levels of autonomy.

In March 2026, Jensen Huang said he would be concerned if an engineer earning \$500{,}000 a year used less than \$250{,}000 worth of AI tokens over the course of that year \citep{allin2026huang}. Token spending adds a usage-based component to the price of delegation. In my setting, a provider could combine an autonomy ceiling with an initial token allowance and a price schedule for additional tokens. Since tokens measure computational input, this extension would require specifying how many tokens the AI uses to complete a given share of a task. When designing its menu, the provider would then anticipate both the user's demand for tokens and the AI's willingness to accept delegation. The question is how upfront prices and usage-based charges jointly affect product design, information rents, and realized delegation.

A user's product choice determines both the technology she receives and what the AI infers about her readiness. Readiness directly raises the value of delegation. As users delegate more work, they gain more from accurate execution. A richer interface can also change which AI types are willing to act, and their acceptance decisions shape the information rents the provider must leave to users. Through these delegation and screening incentives, the optimal menu pairs higher available autonomy with richer tools for guidance and oversight.

%=======================

\clearpage

\section*{Acknowledgments}

I am grateful to the Murphy Institute at Tulane University for its hospitality during my visit. I also thank those whose anonymous comments helped improve the paper. All remaining errors are my own.

\section*{Conflicts of interest}

I have no conflicts of interest.

\section*{Funding}

I received no specific funding for this research.

\section*{Data availability}

The data underlying Figure~\ref{fig:ai_product_scatter} is available in the \hyperref[tab:ai_product_scores]{Appendix}. The source code used to generate the data is available upon request.

%=======================

\clearpage
\bibliographystyle{apalike}
\bibliography{references}

%=============

\clearpage
\appendix

\section*{Appendix}

\section{Proofs}\label{app:proofs}
\renewcommand{\theequation}{A\arabic{equation}}
\renewcommand{\thelemma}{A\arabic{lemma}}
\renewcommand{\theHlemma}{A\arabic{lemma}}
\renewcommand{\theHequation}{A\arabic{equation}}
\setcounter{equation}{0}
\setcounter{lemma}{0}

\begin{lemma}\label{lem:delegation_derivatives}
For an interior AI acceptance cutoff $\theta_A(x,I)=1/(\kappa Ix)$,
\begin{align}
D_x(x,I)
&=
\mu(x,I)
-
\frac{\theta_A(x,I)}{x}f\!\left(\theta_A(x,I)\right),\label{eq:D_readiness_derivative}\\
D_I(x,I)
&=
\frac{1}{\kappa I^2}
\left[
\int_{\theta_A(x,I)}^{\overline\theta}
\frac{f(\theta)}{\theta}\,d\theta
-
\frac{1}{x}f\!\left(\theta_A(x,I)\right)
\right],
\label{eq:D_derivatives}
\end{align}
and
\begin{equation}
\mu_x(x,I)
=
\frac{\theta_A(x,I)}{x}f\!\left(\theta_A(x,I)\right),
\qquad
\mu_I(x,I)
=
\frac{\theta_A(x,I)}{I}f\!\left(\theta_A(x,I)\right).
\label{eq:mu_derivatives}
\end{equation}
\end{lemma}
\begin{proof}[Proof of Lemma~\ref{lem:delegation_derivatives}]
Since
\[
\theta_{A,x}=-\frac{\theta_A}{x},
\qquad
\theta_{A,I}=-\frac{\theta_A}{I},
\qquad
x-\frac{1}{\kappa\theta_A I}-1=-1,
\]
Leibniz's rule gives
\[
D_x=\mu+f(\theta_A)\theta_{A,x},
\qquad
D_I=\frac{1}{\kappa I^2}
\int_{\theta_A}^{\overline\theta}\frac{f(\theta)}{\theta}\,d\theta
+f(\theta_A)\theta_{A,I}.
\]
Substitution yields Equations~\eqref{eq:D_readiness_derivative} and~\eqref{eq:D_derivatives}. Likewise,
$\mu_j=-f(\theta_A)\theta_{A,j}$ for $j\in\{x,I\}$ gives
Equation~\eqref{eq:mu_derivatives}.
\end{proof}

\begin{lemma}\label{lem:screening_envelope}
Consider an incentive-compatible separating menu, possibly with an inactive lower interval, and let $U(x)$ denote equilibrium indirect payoff. Then $U$ is absolutely continuous and, almost everywhere, $U'(x)$ equals the derivative with respect to true readiness of the payoff from the product selected by type $x$. Under the profit-maximizing normalization $U(\underline x)=1$, Equations~\eqref{eq:general_rent} and \eqref{eq:rent_identity} follow. In particular, whenever type $x$ is assigned an active separating product $(I,\bar q,p)$,
\[
U'(x)=\bar q\mu(x,I),
\]
and its local contribution to relaxed profit is $w(x;I,\bar q)g(x)\,dx$, with $w$ given by Equation~\eqref{eq:w}, regardless of the allocation assigned to higher types.
\end{lemma}
\begin{proof}[Proof of Lemma~\ref{lem:screening_envelope}]
Each product payoff in Equation~\eqref{eq:flat_linear_payoff} is the maximum
of a constant and an affine function with slope in $[0,1]$. Their upper
envelope $U$ is therefore non-decreasing and $1$-Lipschitz, hence absolutely
continuous. The incentive-compatibility envelope condition gives
\[
U'(x)=
\begin{cases}
\bar q(x)\mu(x,I(x)),&\text{on the active separating region},\\
0,&\text{on the inactive region},
\end{cases}
\quad\text{almost everywhere}.
\]
If $U(\underline x)>1$, raising all active prices by
$U(\underline x)-1$ preserves product choices and participation. Thus the
profit-maximizing normalization is $U(\underline x)=1$, and
\[
U(x)-1=\int_{\underline x}^{x}U'(r)\,dr,
\qquad
\int_{\underline x}^{\overline x}[U(x)-1]g(x)\,dx
=\int_{\underline x}^{\overline x}U'(r)[1-G(r)]\,dr.
\]
These are Equations~\eqref{eq:general_rent} and \eqref{eq:rent_identity}.
An active separating product reveals $x$, so its contribution after
subtracting information rents is
\[
\bar qD(x,I)-K(I)-C(\bar q)
-\frac{1-G(x)}{g(x)}\bar q\mu(x,I)
=w(x;I,\bar q).
\]
\end{proof}

\begin{lemma}\label{lem:necessary_monotonicity}
For any globally incentive-compatible menu with continuation payoffs of the form in Equation~\eqref{eq:flat_linear_payoff}, the equilibrium indirect payoff $U(x)$ is convex in readiness. Along any differentiable active separating interval, $\bar q(x)\mu(x,I(x))$ is weakly increasing in $x$, so
\[
\frac{d}{dx}\bigl[\bar q(x)\mu(x,I(x))\bigr]\ge 0
\]
wherever the derivative exists.
\end{lemma}
\begin{proof}[Proof of Lemma~\ref{lem:necessary_monotonicity}]
For each contract $\hat x$,
\[
u(x,\hat x)=
\max\left\{
1-p(\hat x),\
1-p(\hat x)+\bar q(\hat x)
\bigl[D(\hat x,I(\hat x))+\mu(\hat x,I(\hat x))(x-\hat x)\bigr]
\right\}
\]
is convex in $x$. Hence $U(x)=\sup_{\hat x}u(x,\hat x)$, including the
baseline, is convex. Lemma~\ref{lem:screening_envelope} gives
$U'=\bar q\mu$ almost everywhere on the active separating region.
Since a convex function has a non-decreasing derivative and $\bar q\mu$
is continuous along a differentiable menu,
\[
\frac{d}{dx}\bigl[\bar q(x)\mu(x,I(x))\bigr]\ge0
\]
wherever the derivative exists.
\end{proof}

\begin{lemma}\label{lem:implementation}
Consider the inactive baseline below $x_0$ and a measurable active allocation on $(x_0,\overline x]$, with $D(x,I(x))>0$. At $x_0$, allow either an active product with $D(x_0,I(x_0))>0$ or the baseline, which gives zero expected delegation. Suppose any common active product induces full AI acceptance for every assigned type, $\bar q(x)\mu(x,I(x))$ is weakly increasing, and prices satisfy
\[
p(x)=\bar q(x)D(x,I(x))-
\int_{x_0}^{x}\bar q(\xi)\mu(\xi,I(\xi))\,d\xi.
\]
The allocation is globally incentive compatible and individually rational, with $U(x_0)=1$, if and only if every active-product price is nonnegative.
\end{lemma}
\begin{proof}[Proof of Lemma~\ref{lem:implementation}]
Set $U(x)=1$ for $x\le x_0$ and
\[
U(x)=1+\int_{x_0}^{x}\bar q(\xi)\mu(\xi,I(\xi))\,d\xi
\]
above it. Monotonicity of $\bar q\mu$ makes $U$ convex. The prescribed price
therefore gives
\[
u(x,\hat x)=
\max\left\{
1-p(\hat x),\
U(\hat x)+\bar q(\hat x)\mu(\hat x,I(\hat x))(x-\hat x)
\right\}
\le U(x)
\]
whenever $p(\hat x)\ge0$: the first term is at most $1$, and the second is
a supporting line to $U$. Since $D(x,I(x))>0$, the assigned product attains
$U(x)\ge1$, proving incentive compatibility and participation.
Conversely, if $p(\hat x)<0$, type $x_0$ can choose that product and turn
delegation off, obtaining $1-p(\hat x)>U(x_0)=1$. This also applies when
$x_0=\underline x$.
\end{proof}

\begin{proof}[\phantomsection\label{proof:global_bound}Proof of Lemma~\ref{lem:global_bound}]
I proceed in three steps.

\noindent \textbf{Step 1. Returns to interface richness and readiness.}
Allow posterior mean readiness $r$ and mean inverse hazard $M$ to vary
separately, and define
\[
B(r,M,I)=D(r,I)-M\mu(r,I),
\quad
W(r,M)=\max\{0,\sup_{I>0,\,0<\bar q\le1}[\bar qB(r,M,I)-K(I)-C(\bar q)]\}.
\]
Thus $B(x,I)=B(x,m(x),I)$ and $W(x)=W(x,m(x))$.
At partial acceptance, write
\[
A=1+M,\quad z=\frac1{\kappa Ir},\quad
J=\int_z^{\overline\theta}\frac{f(\theta)}{\theta}\,d\theta,\quad
a=\frac{zJ}{\mu},\quad \lambda=\frac{f(z)}{\mu}.
\]
By Assumption~\ref{ass:mhr},
$\Pr(\theta-z>t\mid\theta\ge z)\le e^{-\lambda t}$, so
$\E[\theta-z\mid\theta\ge z]\le1/\lambda$. Jensen's inequality yields
\begin{equation}
a=z\E[1/\theta\mid\theta\ge z]\ge\frac{z\lambda}{1+z\lambda}.
\label{eq:reliability_bound}
\end{equation}
If $B=\mu[r-A-ra]>0$, then
\[
\frac rA>\frac1{1-a}\ge1+z\lambda
\ge\frac{z\lambda}{a}=\frac{f(z)}J,
\qquad \frac{Az\lambda}{r}<a<1.
\]
Differentiation gives
\begin{align}
B_I&=\frac zI(rJ-Af(z))>0,
& B_r&=\mu-\frac{Azf(z)}r>0,
\label{eq:general_positive_derivatives}\\
B_{Ir}&=\frac{zf(z)}I
\left[1+\frac Ar\left(1+z\frac{f'(z)}{f(z)}\right)\right]>0,
& B_{rr}&=\frac{zf(z)}{r^2}
\left[r+A\left(2+z\frac{f'(z)}{f(z)}\right)\right]>0.
\label{eq:general_cross_derivatives}
\end{align}
The last two signs follow from $f'/f\ge-\lambda$ and $Az\lambda/r<a<1$.
Also,
\[
B_M=-\mu,\qquad B_{rM}=-\mu_r,\qquad B_{MM}=0,\qquad
B_{IM}=-\frac{zf(z)}I.
\]
Under full acceptance,
$B=r-1-M-\E[1/\theta]/(\kappa I)$.
Thus the positive-$B$ set is an upper interval in $I$.

\noindent \textbf{Step 2. Ordering maximizing products.}
Define
\[
\Psi(B)=\max_{0\le\bar q\le1}\{\bar qB-C(\bar q)\},
\qquad
q(B)=\operatorname*{arg\,max}_{0\le\bar q\le1}
\{\bar qB-C(\bar q)\}.
\]
Strict convexity of $C$ makes $q(B)$ unique, continuous, and non-decreasing,
with strict increase wherever it is interior. Along an ordered line
with constant slopes $dr/dt=\alpha\ge0$ and $dM/dt=-\beta\le0$, take $I_H>I_L$
with $B_L>0$. Step 1 gives $B_H>B_L$ and
$B_{H,t}\ge B_{L,t}>0$ for a nonconstant line. Hence
\[
\frac d{dt}\bigl[\Psi(B_H)-\Psi(B_L)\bigr]
=q(B_H)B_{H,t}-q(B_L)B_{L,t}\ge0.
\]
The cost difference $K(I_H)-K(I_L)$ is constant.
If $B_L\le0$ initially, the lower interface has negative contribution.
At its first positive $B_L$, it still cannot displace a profitable higher
interface; thereafter the displayed inequality applies.
The comparison is strict when the lower product has partial acceptance
or full acceptance with interior autonomy. Two full-acceptance cap
products cannot both maximize because
$K(I)+\E[1/\theta]/(\kappa I)$ is strictly convex.
Since $m$ is nonincreasing, maximizing interfaces are non-decreasing in
$x$. Increasing $B$, and then $\mu$, orders $\bar q$ and $\bar q\mu$.

\noindent \textbf{Step 3. The active interval and attainment.}
A profitable product remains profitable at higher readiness. Thus
$\{x:W(x)>0\}$ is an upper interval. Any positive contribution requires
\[
I>\frac1{\kappa x\overline\theta},\qquad
K(I)<x-1,\qquad
\bar q>\frac{K((\kappa x\overline\theta)^{-1})}{x-1}>0.
\]
Coercivity bounds $I$ above. A maximizing sequence with positive
contribution therefore lies in a compact set bounded away from $I=0$
and $\bar q=0$. Continuity gives attainment.
\end{proof}

\begin{proof}[\phantomsection\label{proof:global_menu}Proof of Theorem~\ref{thm:global_menu}]
I proceed in four steps, using $B(r,M,I)$ and $W(r,M)$ from the proof of
Lemma~\ref{lem:global_bound}.

\noindent \textbf{Step 1. A bound for pooled products.}
For a fixed product, set
$V(r,M)=\max\{0,\allowbreak\bar qB(r,M,I)-K(I)-C(\bar q)\}$.
Along an ordered line of slopes $(\alpha,-\beta)$, $\alpha,\beta\ge0$, its
derivatives where $V>0$ satisfy
\[
V_t=\bar q(\alpha B_r+\beta\mu)\ge0,\qquad
V_{tt}=\bar q[\alpha^2B_{rr}+2\alpha\beta\mu_r]\ge0.
\]
The second inequality is strict under partial acceptance when
$\alpha>0$. Under full acceptance, the untruncated contribution is
affine. Its directional derivative rises weakly at the acceptance
boundary and at entry from zero. Thus $V$, and its supremum $W$, are
convex along every ordered line.

Since $m$ is nonincreasing, splitting a finite posterior into lower and
upper readiness groups gives the upper group a weakly lower mean inverse
hazard. The two group means therefore lie on an ordered line. Applying
convexity to the split and iterating gives
\begin{equation}
W(\E X,\E m(X))\le\E W(X,m(X)).
\label{eq:posterior_bound}
\end{equation}
Finite approximations extend this inequality to arbitrary posteriors;
$0\le W(r,M)\le r-1$ and
$\int m(x)\,dG(x)=\int[1-G(x)]\,dx<\infty$ justify the limits.

\noindent \textbf{Step 2. The global profit bound.}
For a product with acceptance probability $\mu$, let
\[
L=\frac1{\kappa I}\int_{\theta_A}^{\overline\theta}
\frac{f(\theta)}{\theta}\,d\theta.
\]
Its continuation payoff is $1-p+\bar q[(x-1)\mu-L]_+$.
The upper-envelope argument in Lemma~\ref{lem:screening_envelope} gives
$U'=\bar q\mu$ for activating users and $U'=0$ otherwise.
After subtracting information rents, an activating user contributes
\[
\bar q[(x-1-m(x))\mu-L]-K(I)-C(\bar q).
\]
For that product's activating purchasers, set
$r=\E[X\mid\text{product, activation}]$ and
$M=\E[m(X)\mid\text{product, activation}]$.
By Equation~\eqref{eq:ai_pooled_acceptance}, the AI uses the mean readiness
$r$ of precisely these purchasers. Their mean contribution is $\bar qB(r,M,I)-K(I)-C(\bar q)$.
A nonactivating purchaser pays zero; omitting her nonnegative provision
cost raises the bound. Equation~\eqref{eq:posterior_bound} and
$U(\underline x)\ge1$ therefore imply
\[
\Pi\le\int_{\underline x}^{\overline x}W(x,m(x))\,dG(x),
\]
which is Equation~\eqref{eq:global_profit_bound}, including for menus
with infinitely many products.

\noindent \textbf{Step 3. Implementation and attainment.}
Along the selected maximizers, $B$ strictly increases with readiness.
Interior autonomy therefore strictly increases. If autonomy is capped
while acceptance is partial, $B_{Ix}>0$ rules out a common maximizing
interface for two distinct types. The only possible common product is
the unique full-acceptance cap product. Ordering makes it terminal, and
full acceptance makes its posterior consistent.

Positive virtual profit gives $D\ge B>0$.
Lemma~\ref{lem:global_bound} gives non-decreasing effective delegation.
Equation~\eqref{eq:global_price} and $p^\ast\ge0$ then satisfy
Lemma~\ref{lem:implementation}, yielding global incentive compatibility
and participation. Each selected product attains $W(x)$, so the menu
attains the bound in Step 2.

\noindent \textbf{Step 4. Pooling in any optimum.}
Equality in Step 2 is necessary for every optimum. A negative mean
product contribution gives a strict inequality relative to the baseline.
For a nondegenerate active posterior with partial acceptance, Step 1
also gives strict inequality: if the mean contribution is positive,
$V_{tt}>0$ near the mean; if it is zero, positive provision costs imply
$B>0$ and $V$ has a strict kink there. Thus an optimal active pool has
full acceptance at its posterior mean.

For its fixed product, define the affine full-acceptance contribution
\[
\Lambda(r,M)=\bar q\left[r-1-M-\frac{\E[1/\theta]}{\kappa I}\right]
-K(I)-C(\bar q).
\]
Rejecting types below $1/(\kappa Ir)$ removes strictly negative virtual
surplus, so actual contribution is at least $\Lambda(r,M)$, strictly so under
partial acceptance. At the fully accepting posterior mean,
\[
\E W(X,m(X))
\ge \E w(X;I,\bar q)
\ge \E\Lambda(X,m(X))
=\Lambda(r,M).
\]
Equality requires full acceptance and pointwise optimality at almost
every underlying readiness type. If $\bar q<1$, this would require
\[
C'(\bar q)=x-1-m(x)-\frac{\E[1/\theta]}{\kappa I}
\]
for a positive mass of distinct types, although the right side is
strictly increasing in $x$. Hence $\bar q=1$. Strict convexity gives the
unique interface minimizing $K(I)+\E[1/\theta]/(\kappa I)$; ordering makes
the pool terminal, and the envelope price is constant on it.
\end{proof}

\begin{proof}[\phantomsection\label{proof:positive_prices}Proof of Corollary~\ref{cor:positive_prices}]
Suppose the active interval is nonempty, with entry boundary
$x_0<\overline x$. The compactness bounds in Lemma~\ref{lem:global_bound}
give an active limiting maximizer with interface $I_0$, including when
$W(x_0)=0$, with $B(x_0,I_0)>0$. The limiting $m_0=m(x_0)$ is finite,
since unbounded $m$ precludes positive $B$ near entry.

For an active $x\ge x_0$, fix $I=I^\ast(x)\ge I_0$ and write
$\mu_0=\mu(x_0,I)$. Then $D(x_0,I)>m_0\mu_0$. Under Assumption~\ref{ass:mhr},
\[
\partial_r\log\mu(r,I)=\frac{\theta_A(r,I) f(\theta_A(r,I))}{r\mu(r,I)}
\]
is nonincreasing in $r$ and falls to zero at full acceptance.
Equation~\eqref{eq:reliability_bound} bounds its entry value by $1/(1+m_0)$.
Since $g$ is non-decreasing,
\[
m_0=\int_{x_0}^{\overline x}\frac{g(t)}{g(x_0)}\,dt
\ge\overline x-x_0\ge x-x_0,
\]
and therefore
\[
\frac{\mu(x,I)}{\mu_0}
\le e^{(x-x_0)/(1+m_0)}
\le e^{m_0/(1+m_0)}<1+m_0.
\]
The inequality also holds if acceptance is full at entry. Using
$D_r=\mu-\mu_r$ and the ordering of $I^\ast$ and $\bar q^\ast$ gives
\begin{align*}
p^\ast(x)
&\ge\bar q^\ast(x)
\left[D(x,I)-\int_{x_0}^{x}\mu(t,I)\,dt\right]\\
&=\bar q^\ast(x)[D(x_0,I)+\mu_0-\mu(x,I)]>0.
\end{align*}
\end{proof}

\begin{proof}[\phantomsection\label{proof:continuous_menu}Proof of Theorem~\ref{thm:continuous_menu}]
I proceed in three steps. Fix readiness $x$, and set
$a=(1+m(x))/x$, $t=\log I$, and $z=1/(\kappa Ix)$.
Positive optimal profit requires $x-1-m(x)>C'(0+)\ge0$, so $0<a<1$.

\noindent \textbf{Step 1. Shape of the interface return.}
On the positive-$B$ partial-acceptance region, define
\[
J(z)=\int_z^{\overline\theta}\frac{f(\theta)}{\theta}\,d\theta,
\qquad N(z)=J(z)-af(z).
\]
The proof of Lemma~\ref{lem:global_bound} gives $B_t=xzN>0$.
With $y=\log z$,
\[
N_y=-f-azf'<0,\qquad
N_{yy}=-z[(1+a)f'+azf'']\le0,
\]
because the density conditions imply $f'\ge0$, $2f'+zf''\ge0$, and
\[
(1+a)f'+azf''=(1-a)f'+a(2f'+zf'')\ge0.
\]
Thus $\log B_t=\text{constant}-t+\log N$ is strictly concave.
At $z=\underline\theta$, $f(\underline\theta+)=0$ makes $B_t$ continuous,
and the left and right derivatives of $\log B_t$ are, respectively,
\[
-1+\frac{a\underline\theta f'(\underline\theta+)}
{J(\underline\theta)}
\quad\text{and}\quad -1.
\]
The limit of $f'$ is finite because $\theta^2f'$ is nonnegative and
non-decreasing. Hence $B_t$ remains log-concave across the boundary;
under full acceptance, $B_t=\E[1/\theta]/(\kappa I)$.

Where $B_t$ is differentiable, write $k=(\log B_t)'$. Integrating the tangent bound for $\log B_t$ from the lower endpoint where $B=0$ gives $B<B_t/k$ if $k>0$. For $k\le0$, $B_t/B>k$ as well. Therefore
\[
(\log B)''=\frac{B_t}{B}\left(k-\frac{B_t}{B}\right)<0.
\]
The derivative $(\log B)'=B_t/B$ is continuous at full acceptance, so strict
concavity extends across the boundary.

\noindent \textbf{Step 2. Uniqueness of the positive maximizer.}
Using the autonomy choice $\bar q=q(B)$ from Lemma~\ref{lem:global_bound},
\[
\frac{d\log q(B)}{d\log B}=\frac{C'(\bar q)}{\bar q C''(\bar q)}
\]
where autonomy is interior, and the derivative falls to zero at the cap.
Thus $\log q(e^v)$ is concave and non-decreasing. Step 1 makes
$\log q(B)$ concave in $t$. Also, $\log[IK'(I)]$ is convex because its
derivative in $t$ is $1+IK''(I)/K'(I)$.

Define
\[
R(t)=\frac{q(B)B_t}{IK'(I)}.
\]
Then $\log R$ is concave and
\[
\frac{d}{dt}\bigl[\Psi(B(x,e^t))-K(e^t)\bigr]=IK'(I)[R(t)-1].
\]
At the lower positive-autonomy endpoint, $q(B)\to0$ while $I>0$,
so $R\to0$. At infinity, $B_t=\E[1/\theta]/(\kappa I)$ and $IK'$ grows,
so again $R\to0$. Hence $\{t:R(t)>1\}$ is an interval. The profile starts
below zero and tends to $-\infty$, so a positive global maximum occurs
only at the final downward crossing of $R=1$. Concavity of $\log R$
rules out a flat maximizing interval after $R>1$. The positive maximizer
is therefore unique.

\noindent \textbf{Step 3. Continuity and implementation.}
Positive profit locally bounds $I$ and $\bar q$ away from zero, and
coercivity bounds $I$ above. The maximum theorem gives continuity of the
unique active maximizer, including at full acceptance and the autonomy
cap. Corollary~\ref{cor:positive_prices} gives positive prices and global
implementation. Equality in the bound of Theorem~\ref{thm:global_menu},
together with uniqueness, makes every optimum agree with this allocation
almost everywhere on the positive-profit region.

Now suppose that a positive-autonomy product ties the baseline at an $x$
with $W(x)=0$. Its provision cost is positive, so $B(x,I)>0$. Step~1
of the proof of Lemma~\ref{lem:global_bound} shows that, at this fixed
interface, $B$ then rises strictly with readiness. The same product
therefore earns positive virtual profit at every higher readiness.
At most one type with $W(x)=0$ can have such a tie, and $G$ assigns
that type probability zero. Buying an active product without activating
it yields zero revenue and positive provision costs. Equality in the profit
bound therefore rules out such purchases on a set of positive probability. Thus the baseline is the unique allocation almost everywhere
on $\{W=0\}$, which gives uniqueness of the implemented
allocation almost everywhere on the full support.
\end{proof}

\begin{proof}[\phantomsection\label{proof:acceptance_boundary}Proof of Proposition~\ref{prop:acceptance_boundary}]
At $\underline I=1/(\kappa x\underline\theta)$,
\[
B_I(x,\underline I-)
=\frac1{\kappa\underline I^2}
\left[\E[1/\theta]-\frac{1+m(x)}x f(\underline\theta+)\right],
\qquad
B_I(x,\underline I+)
=\frac{\E[1/\theta]}{\kappa\underline I^2}.
\]
Since $1/(\kappa\underline I x)=\underline\theta$, subtraction gives
Equation~\eqref{eq:acceptance_jump}. For
$P(I)=\Psi(B(x,I))-K(I)$, optimizing autonomy gives
\[
P'(\underline I+)-P'(\underline I-)
=q(B)\bigl[B_I(x,\underline I+)-B_I(x,\underline I-)\bigr]>0
\]
at positive autonomy. This contradicts the necessary conditions
$P'_-\ge0\ge P'_+$ for a local maximum. A continuous active allocation
crossing from partial to full acceptance would attain this excluded
boundary product.
\end{proof}

\begin{proof}[\phantomsection\label{proof:regular_sorting}Proof of Proposition~\ref{prop:regular_sorting}]
\setlength{\emergencystretch}{2em}
At an interior stationary product,
$B=C'(\bar q)>0$ and $B_I=K'(I)/\bar q>0$.
Equations~\eqref{eq:general_positive_derivatives} and
\eqref{eq:general_cross_derivatives}, with $r=x$ and $M=m(x)$, give
\[
B_x=B_r-m'(x)\mu>0,\qquad
B_{Ix}=B_{Ir}-m'(x)\mu_I>0.
\]
Differentiating the first-order conditions yields
\[
-\begin{pmatrix}
w_{II}&w_{I\bar q}\\
w_{I\bar q}&w_{\bar q\bar q}
\end{pmatrix}
\begin{pmatrix}I^{*\prime}\\\bar q^{*\prime}\end{pmatrix}
=\begin{pmatrix}\bar q^\ast B_{Ix}\\B_x\end{pmatrix}.
\]
The matrix is positive definite with negative off-diagonal entries
$-B_I$, so its inverse has strictly positive entries.
Thus $I^{*\prime},\bar q^{*\prime}>0$.
Under partial acceptance, $\mu_x,\mu_I>0$, so
\[
W'=\bar q^\ast B_x>0,\qquad
Q^{*\prime}=(\bar q^\ast\mu)'=
\mu\bar q^{*\prime}
+\bar q^\ast(\mu_x+\mu_I I^{*\prime})>0.
\]
\end{proof}

\begin{proof}[\phantomsection\label{proof:precision_separation}Proof of Corollary~\ref{cor:precision_separation}]
I proceed in two steps. Since $D$ and $\mu$ depend on $\kappa$ and $I$
through $\kappa I$,
\[
D_\kappa=\frac I\kappa D_I,\qquad
\mu_\kappa=\frac I\kappa\mu_I.
\]

\noindent \textbf{Step 1. Active products.}
Differentiating the first-order conditions and using
$K'(I^\ast)=\bar q^\ast B_I$ gives
\[
-\begin{pmatrix}
w_{II}&w_{I\bar q}\\
w_{I\bar q}&w_{\bar q\bar q}
\end{pmatrix}
\begin{pmatrix}
\displaystyle\frac{\partial I^\ast}{\partial\kappa}
+\frac{I^\ast}{\kappa}\\[6pt]
\displaystyle\frac{\partial\bar q^\ast}{\partial\kappa}
\end{pmatrix}
=
\begin{pmatrix}
\displaystyle\frac{K'(I^\ast)+I^\ast K''(I^\ast)}{\kappa}\\[6pt]
0
\end{pmatrix}.
\]
The inverse matrix has strictly positive entries, as in
Proposition~\ref{prop:regular_sorting}. Hence
\[
\frac{\partial\bar q^\ast}{\partial\kappa}>0,\qquad
\frac{\partial I^\ast}{\partial\kappa}+\frac{I^\ast}{\kappa}>0,
\qquad
\frac{\partial(\kappa I^\ast)}{\partial\kappa}>0,
\]
and
\[
\frac{\partial Q^\ast}{\partial\kappa}
=\mu\frac{\partial\bar q^\ast}{\partial\kappa}
+\bar q^\ast\mu_I
\left(\frac{\partial I^\ast}{\partial\kappa}
+\frac{I^\ast}{\kappa}\right)>0.
\]

\noindent \textbf{Step 2. Entry.}
The envelope theorem and Equation~\eqref{eq:foc_I} imply
\[
\partial_\kappa W
=\bar q^\ast\frac{I^\ast}{\kappa}B_I
=\frac{I^\ast K'(I^\ast)}{\kappa}>0.
\]
At a locally movable interior entry boundary, differentiate the
zero-profit condition along the smooth extension of the active branch.
Since $W_x>0$ by Proposition~\ref{prop:regular_sorting},
\[
\frac{dx_0}{d\kappa}
=-\frac{\partial_\kappa W(x_0(\kappa)+;\kappa)}
{\partial_x W(x_0(\kappa)+;\kappa)}<0.
\]
\end{proof}

\begin{proof}[\phantomsection\label{proof:high_precision}Proof of Proposition~\ref{prop:high_precision}]
I proceed in five steps.

\noindent \textbf{Step 1. A sufficient precision condition.}
Set
\[
K_0=\lim_{I\downarrow0}K(I),\qquad
b=\min\left\{\frac{\underline x\underline\theta}{\overline\theta},
\underline x-1\right\}>0,
\]
and define
\[
H(v)=\max\left\{0,\sup_{0<s\le1}[sv-C(s)-K_0]\right\},
\qquad
\gamma=\inf_{0<s\le1}\frac{C(s)+K_0}{s}.
\]
The function $H$ is convex and non-decreasing, with
$H(\gamma)=0$ and $H(\gamma+b/2)>0$.
A sufficient condition is
\begin{equation}
K\!\left(\frac{2\E[1/\theta]}{\kappa b}\right)-K_0
<H(\gamma+b/2).
\label{eq:high_precision_condition}
\end{equation}
Its left side decreases to zero as $\kappa\to\infty$, so the condition
holds for all sufficiently large $\kappa$.

\noindent \textbf{Step 2. Profit in an arbitrary menu.}
For any product inducing positive delegation, let $\theta_A$ and $\mu$
be its actual acceptance cutoff and probability. The delegation-weighted
mean readiness entering $\theta_A$ lies in
$[\underline x,\overline x]$. Define
\[
s=\bar q\mu,\qquad
\ell=\frac1{\kappa I\mu}
\int_{\theta_A}^{\overline\theta}\frac{f(\theta)}{\theta}\,d\theta.
\]
Holding its equilibrium posterior fixed, the product gives payoff
$1-p+[s(x-1-\ell)]_+$. For a user of readiness $x$, let $s(x)$ equal
$\bar q\mu$ if she activates her chosen product and zero otherwise. The
upper envelope over products and the baseline is convex and Lipschitz.
Thus $s(x)$ is non-decreasing,
$U'(x)=s(x)$ almost everywhere, and $U(\underline x)\ge1$.
A nondelegating purchaser pays zero. Dropping her nonnegative costs and
integrating information rents gives
\begin{equation}
\begin{split}
\Pi\le{}&
\int_{\underline x}^{\overline x}
\left[s(x)\phi(x)-s(x)\ell(x)-K(I(x))-C(\bar q(x))\right]dG(x)\\
&-\left[U(\underline x)-1\right],
\end{split}
\label{eq:high_precision_menu_bound}
\end{equation}
with zero integrand for nondelegating users. This includes products with
$\mu=0$. At a virtual coefficient $v$, write
$j(v)=s(v-\ell)-K(I)-C(\bar q)$.

\noindent \textbf{Step 3. Bounding execution losses.}
For every product with $\mu>0$,
\[
\ell\ge\frac1{\kappa I\overline\theta},
\qquad
\ell<b\ \Longrightarrow\
\frac1{\kappa I}<b\overline\theta
\le\underline x\underline\theta.
\]
Thus $\ell<b$ guarantees full AI acceptance at every posterior mean and
positive activation gains for every user, since $b\le\underline x-1$.
Partial acceptance therefore requires $\ell\ge b$.

For $\ell\ge b$, $s\le\bar q$, monotonicity of $C$, and $K(I)\ge K_0$
give
\[
j(v)\le s(v-b)-C(s)-K_0\le H(v-b).
\]
If $j(v)<0$, the baseline strictly dominates. If $j(v)\ge0$, then
$v-b\ge\gamma$. Choose
\[
J=\frac{2\E[1/\theta]}{\kappa b}.
\]
Since $\E[1/\theta]\ge1/\overline\theta$,
\[
\frac1{\kappa J}
=\frac{b}{2\E[1/\theta]}
\le\frac{b\overline\theta}{2}
\le\frac{\underline x\underline\theta}{2}.
\]
This interface guarantees full acceptance and loss $b/2$.
Optimizing its autonomy gives
$H(v-b/2)-[K(J)-K_0]$, with positive autonomy.
Convexity of $H$ and Equation~\eqref{eq:high_precision_condition} imply
\[
H(v-b/2)-H(v-b)
\ge H(\gamma+b/2)-H(\gamma)>K(J)-K_0.
\]
Hence every product with $\ell\ge b$ is strictly dominated in virtual
contribution. The same comparison applies to formal full-acceptance
candidates, with $s=\bar q$ and $\ell=\E[1/\theta]/(\kappa I)$.

\noindent \textbf{Step 4. The reduced problem and ironing.}
For $\bar q>0$, $K(I)+\bar q\E[1/\theta]/(\kappa I)$ is strictly convex
and diverges at both ends of its domain. Its unique minimizer satisfies
Equation~\eqref{eq:high_precision_interface} and is continuous in $\bar q$.
Moreover,
\[
K_0\le c_\kappa(\bar q)
\le C(\bar q)+K(\sqrt{\bar q})
+\frac{\E[1/\theta]\sqrt{\bar q}}{\kappa}
\longrightarrow K_0
\quad\text{as }\bar q\downarrow0.
\]
Since $c_\kappa(0)=0$, the cost is lower semicontinuous on $[0,1]$, and
\[
W_\kappa(v)=\max_{0\le\bar q\le1}
\{\bar qv-c_\kappa(\bar q)\}
\]
attains its maximum. Step 3 excludes positive-autonomy maximizers with
loss at least $b$. Every active maximizer therefore induces full
acceptance at every posterior mean and positive activation gains for
every user. For every actual competing product,
$j(v)\le W_\kappa(v)$, strictly when $\ell\ge b$.

The quantile construction in Equation~\eqref{eq:ironed_readiness} gives
$\Phi-\bar\Phi\ge0$, with equality at both endpoints. Integration by parts
against non-decreasing $s$ yields
\[
\int_{\underline x}^{\overline x}
s(x)[\phi(x)-\bar\phi(x)]\,dG(x)
=-\int_0^1[\Phi(z)-\bar\Phi(z)]\,d[s(G^{-1}(z))]
\le0.
\]
Equality holds when $s$ is constant on each ironing interval.
Combining this inequality with Equation~\eqref{eq:high_precision_menu_bound}
and the product bound gives
\begin{equation}
\Pi\le\int_{\underline x}^{\overline x}W_\kappa(\bar\phi(x))\,dG(x).
\label{eq:high_precision_upper_bound}
\end{equation}

\noindent \textbf{Step 5. Attainment and global incentives.}
For $v_2>v_1$ and maximizing choices $\bar q_2,\bar q_1$, adding their
optimality inequalities gives
\[
(\bar q_2-\bar q_1)(v_2-v_1)\ge0.
\]
Choose a common maximizer at each $\bar\phi$ and the baseline whenever
the maximum is zero. This gives non-decreasing $\bar q_\kappa$, constant
on ironing intervals; define endpoint values by monotone limits.
Equation~\eqref{eq:high_precision_interface} makes
$I_\kappa(\bar q)$ strictly increasing in $\bar q$, since $I^2K'(I)$ is
strictly increasing.

Set
\[
U(x)=1+\int_{\underline x}^{x}\bar q_\kappa(r)\,dr.
\]
Monotonicity and Equation~\eqref{eq:high_precision_price} give
\[
U(x)-1\le\bar q_\kappa(x)(x-\underline x),\qquad
p_\kappa(x)\ge\bar q_\kappa(x)
\left[\underline x-1-
\frac{\E[1/\theta]}{\kappa I_\kappa(\bar q_\kappa(x))}\right]>0.
\]
For any active product $\hat x$, its payoff is
\[
u(x,\hat x)=
\max\{1-p_\kappa(\hat x),\
U(\hat x)+\bar q_\kappa(\hat x)(x-\hat x)\}
\le U(x),
\]
because $U$ is convex. The assigned product attains $U(x)\ge1$, proving
global incentive compatibility and participation.
On an interval with constant $\bar q_\kappa$, the interface is constant
and the envelope formula makes the price constant. Full acceptance for
every posterior mean ensures consistency of the resulting pools.

The allocation attains $W_\kappa$ pointwise, is constant on ironing
intervals, and has $U(\underline x)=1$. It therefore attains
Equation~\eqref{eq:high_precision_upper_bound}, proving
Equation~\eqref{eq:high_precision_profit}.
Any optimum with partial acceptance for a positive mass of active users
would make the product bound strict. Purchases of active products with zero delegation
yield zero revenue and strictly positive provision costs. Hence every
optimum has full AI acceptance for almost every active user type.
\end{proof}

\pagebreak

\section{Coding of Figure~\ref{fig:ai_product_scatter} in the main paper}\label{app:coding}

\renewcommand{\thetable}{B\arabic{table}}
\setcounter{table}{0}

Figure~\ref{fig:ai_product_scatter} in the main paper maps a selected set of AI examples, including commercial products, prototypes, and configured or generic deployments. I use documentation collected on May 7 and 8, 2026, and AI-assisted coding to construct descriptive indices of interface richness and available autonomy. In particular, the autonomy index records capabilities that a configuration permits or supports. Actual delegation also depends on whether the AI offers to continue and how much work the user delegates after that offer.

I selected both named products and broader deployment categories to illustrate variation in product design. I assign each entry to one display category solely to label it in the figure. OSS denotes open-source software and LLM denotes a large language model.

\begin{table}[H]
\centering
\small
\caption{Observation categories}\label{tab:products}
\begin{tabular}{ll}
\toprule
No. & Category \\
\midrule
1 & For-profit general \\
2 & For-profit specialized \\
3 & Internal deployment \\
4 & Research prototype \\
5 & OSS prototype \\
\bottomrule
\end{tabular}
\end{table}

I used OpenAI's GPT model 5.5 to code each entry on eight available-autonomy features and eight interface-richness features. I chose the features to distinguish between \emph{independent action that the product permits or supports} and \emph{the structure through which a user specifies and governs that action}. In the tables, $\bar q$ labels the score for independent task execution, and $I$ labels the score for the tools users have to specify, guide, and oversee that work. Some product features contribute to both interface richness and available autonomy. The resulting association therefore reflects both the documented product configurations and the choice of feature definitions.

The selected interface-richness features are:

\begin{table}[H]
\centering
\small
\caption{Interface richness ($I$) feature components}\label{tab:Ifeatures}
\begin{tabular}{llll}
\toprule
No. & Feature & No. & Feature \\
\midrule
1 & Structure & 5 & Intervention \\
2 & Context & 6 & Reusability \\
3 & Governance & 7 & Deployment \\
4 & Observability & 8 & Collaboration \\
\bottomrule
\end{tabular}
\end{table}

The selected available-autonomy features are:

\begin{table}[H]
\centering
\small
\caption{Available-autonomy ($\bar q$) feature components}\label{tab:Qfeatures}
\begin{tabular}{llll}
\toprule
No. & Feature & No. & Feature \\
\midrule
1 & Planning & 5 & Workflow coordination \\
2 & Information access & 6 & Iteration \\
3 & Tool integration & 7 & Workflow execution \\
4 & Artifact creation & 8 & Background continuation \\
\bottomrule
\end{tabular}
\end{table}

GPT assigned each feature a score on the unit interval $[0,1]$ using the archived documentation. A score of zero means the documentation provided no evidence that the configuration supports the feature. A score of one denotes a central and explicitly documented feature. Intermediate scores reflect how fully the configuration supports the feature, including any limits on its use described in the documentation. For each index, I average its eight component scores.

Table~\ref{tab:ai_product_scores} reports the coordinates used in Figure~\ref{fig:ai_product_scatter}. In the replication package, \path{ai_product_coded_data.csv} lists the component scores and supporting evidence by configuration. The source manifest identifies the dated records in \path{documentation/}, and \path{README.md} describes the coding procedure.

\begin{table}[htbp]
\centering
\footnotesize
\setlength{\tabcolsep}{3.5pt}
\renewcommand{\arraystretch}{0.88}
\caption{Coding used for Figure~\ref{fig:ai_product_scatter} in the main paper}
\label{tab:ai_product_scores}
\begin{tabular}{rl>{\raggedright\arraybackslash}p{0.36\textwidth}cc}
\toprule
No. & Category & Product & Available autonomy ($\bar q$) & Interface ($I$) \\
\midrule
1 & For-profit general & ChatGPT agent & 0.821 & 0.706 \\
2 & For-profit general & ChatGPT chat & 0.325 & 0.288 \\
3 & For-profit general & Claude Code & 0.900 & 0.811 \\
4 & For-profit general & Claude chat & 0.312 & 0.286 \\
5 & For-profit general & Codex & 0.906 & 0.840 \\
6 & For-profit general & Cursor agent & 0.881 & 0.806 \\
7 & For-profit general & Devin & 0.929 & 0.775 \\
8 & For-profit general & Gemini chat & 0.223 & 0.166 \\
9 & For-profit general & GitHub Copilot coding agent & 0.869 & 0.725 \\
10 & For-profit general & Grok & 0.381 & 0.244 \\
11 & For-profit general & Lovable & 0.575 & 0.806 \\
12 & For-profit general & Microsoft 365 Copilot & 0.494 & 0.656 \\
13 & For-profit general & Replit Agent & 0.731 & 0.788 \\
14 & For-profit general & Windsurf & 0.800 & 0.731 \\
15 & For-profit specialized & Educational scaffolding AI & 0.406 & 0.762 \\
16 & For-profit specialized & Enterprise compliance copilots & 0.425 & 0.850 \\
17 & For-profit specialized & Harvey legal assistant & 0.578 & 0.781 \\
18 & For-profit specialized & Medical documentation AI & 0.488 & 0.838 \\
19 & For-profit specialized & Microsoft Copilot Studio agents & 0.816 & 0.902 \\
20 & For-profit specialized & Salesforce Agentforce & 0.796 & 0.931 \\
21 & For-profit specialized & Underwriting/review assistants & 0.694 & 0.869 \\
22 & Internal deployment & Enterprise governed agents & 0.819 & 0.900 \\
23 & Internal deployment & Internal scripts with LLM tool use & 0.641 & 0.462 \\
24 & Internal deployment & Procurement copilots & 0.672 & 0.713 \\
25 & Internal deployment & Robotic process automation (RPA) with LLM worker & 0.873 & 0.906 \\
26 & Internal deployment & Template-based medical copilots & 0.338 & 0.778 \\
27 & OSS prototype & AutoGPT & 0.784 & 0.519 \\
28 & OSS prototype & Browser Use & 0.713 & 0.431 \\
29 & OSS prototype & CAMEL & 0.512 & 0.475 \\
30 & OSS prototype & Local open-source chat & 0.101 & 0.046 \\
31 & OSS prototype & Open Interpreter & 0.693 & 0.250 \\
32 & OSS prototype & OpenHands & 0.906 & 0.819 \\
33 & Research prototype & BabyAGI & 0.613 & 0.138 \\
34 & Research prototype & Generative Agents / Smallville & 0.662 & 0.412 \\
35 & Research prototype & Voyager & 0.850 & 0.325 \\
\bottomrule
\end{tabular}
\end{table}

\FloatBarrier

\section{Examples of continuous menus}\label{oa:continuity}

\subsection{An example satisfying the continuity conditions}

Let readiness be uniform on $[\underline x,\overline x]$, with $1<\underline x<\overline x$. Then
\[
g(x)=\frac{1}{\overline x-\underline x},
\qquad
m(x)=\frac{1-G(x)}{g(x)}=\overline x-x.
\]
The readiness density is constant and its hazard rate, $1/(\overline x-x)$, is increasing.

Let reliability have a shifted triangular distribution on $[\underline\theta,\overline\theta]$, with $0<\underline\theta<\overline\theta$:
\[
F(\theta)=\frac{(\theta-\underline\theta)^2}{(\overline\theta-\underline\theta)^2},
\qquad
f(\theta)=\frac{2(\theta-\underline\theta)}{(\overline\theta-\underline\theta)^2}.
\]
This density is positive and twice continuously differentiable on the interior of its support. Moreover,
\[
f(\underline\theta+)=0,
\qquad
\frac{d}{d\theta}\bigl[\theta^2f'(\theta)\bigr]
=\frac{4\theta}{(\overline\theta-\underline\theta)^2}>0.
\]
Its hazard rate is increasing because $f$ is increasing and $1-F$ is decreasing. Finally, take quadratic costs $K(I)=kI^2$ and $C(\bar q)=c\bar q^2$, with $k,c>0$. Both costs are strictly increasing and strictly convex on their positive domains, and both marginal-cost elasticities equal one.
Thus all conditions in Assumption~\ref{ass:continuity} of the main paper hold.

The acceptance rule also illustrates the role of the lower-end density condition. With $\theta_A(x,I)=1/(\kappa Ix)$, partial acceptance gives
\[
\mu(x,I)
=1-\frac{[\theta_A(x,I)-\underline\theta]^2}{(\overline\theta-\underline\theta)^2},
\qquad
\mu_I(x,I)
=\frac{2[\theta_A(x,I)-\underline\theta]\theta_A(x,I)}{I(\overline\theta-\underline\theta)^2}.
\]
As $I$ increases to $1/(\kappa x\underline\theta)$, the acceptance share approaches one and $\mu_I$ approaches zero. Beyond this interface level, all AI types accept and $\mu_I=0$. The vanishing density at the lower bound therefore removes the jump in marginal interface returns identified in Proposition~\ref{prop:acceptance_boundary} of the main paper.

Theorem~\ref{thm:continuous_menu} in the main paper gives a unique maximizing product at every readiness with $W(x)>0$, varying continuously with readiness. The parameters $k$, $c$, and $\kappa$ determine which users receive active products and whether the allocation reaches full acceptance or the autonomy cap.

\subsection{Uniform reliability}\label{app:uniform}

\begin{example}[A continuous menu with uniform reliability]\label{ex:uniform_example}
Let readiness be uniform on $[10,11]$, reliability be uniform on $[1/100,1]$, and let $\kappa=1$, $C(\bar q)=10\bar q^2$, and $K(I)=kI^2$. For every $1/400\le k\le1/3$, every readiness type has a unique smooth globally optimal product. AI acceptance is strictly partial and available autonomy is strictly interior throughout the support. The menu has no active jumps or pools.
\end{example}

\begin{proof}[Proof of Example~\ref{ex:uniform_example}]
Here $m(x)=11-x$, $\phi(x)=x-1-m(x)=2x-12\in[8,10]$, $\E[1/\theta]=100\log(100)/99<6$, and $\underline I=100/x>9$. At full acceptance $B=\phi(x)-\E[1/\theta]/I$, and $7<B(x,{\underline I})<10$. Since $B<\phi(x)\le10$, optimal autonomy is always $\bar q=B/20<1/2$ when positive. Where $B>0$, the profiled profit derivative is
\[
\frac{d}{dI}\left[\frac{B^2}{40}-kI^2\right]=2I[k_{\mathrm{req}}(I)-k],
\qquad k_{\mathrm{req}}(I)=\frac{B B_I}{40I}.
\]
In the full-acceptance region,
\[
k_{\mathrm{req}}(I)=\frac{\E[1/\theta](\phi(x)-\E[1/\theta]/I)}{40I^3}.
\]
It is strictly decreasing there because $4\E[1/\theta]-3\phi(x)I<24-216<0$. Its maximum therefore occurs at the boundary and is less than $10\cdot6/(40\cdot9^3)=1/486<1/400\le k$. Thus the full-acceptance profile is strictly decreasing.

In the positive partial region put $h=(1+m(x))/x$, $N=\log(Ix)-h$, and $Y=(99/100)I B=(1-h)e^{N+h}-N-1>0$. Here Equation~\eqref{eq:general_positive_derivatives} of the main paper gives $B_I=100N/(99I^2)>0$, so $N>0$. Since $dY/dN=Y+N$,
\[
\frac{d\log k_{\mathrm{req}}}{d\log I}
=\frac1N+\frac N Y-3,
\]
and its derivative with respect to $N$ is
\[
-\frac1{N^2}+\frac{1-N}{Y}-\frac{N^2}{Y^2}<0.
\]
For $N\ge1$ the sign is immediate. For $0<N<1$, use $(1-h)e^h\le1$ to obtain $Y\le e^N-N-1<N^2/(1-N)$; the last two terms then have a negative sum. Hence the partial marginal-return coefficient increases and then decreases, with at most one positive local maximum of the profit profile. It starts at zero when $B$ first becomes positive. The upward derivative jump at full acceptance excludes the boundary as a maximum.

At the strictly partial test interface $I=1$,
\[
B(x,1)=\frac{\phi(x)-(\phi(x)/x+\log x)}{99/100}>4,
\]
using $\phi(x)\ge8$, $\phi(x)/x<1$, and $\log x<3$. Its optimized virtual profit therefore exceeds $2/5-k\ge1/15$. Every type has a positive unique partial maximum, lying on the strictly decreasing part of $k_{\mathrm{req}}$. This maximum is nondegenerate, so the implicit-function theorem makes the allocation smooth. Corollary~\ref{cor:positive_prices} in the main paper gives positive prices and global implementation. Strict sorting rules out pools.
\end{proof}

\section{Why the first-order conditions are not sufficient}\label{oa:selection}

The product first-order conditions can have several solutions even when the optimal product is unique. Under the monotone-hazard conditions in Assumption~\ref{ass:mhr} of the main paper, I show that whenever a regular maximizing product has partial AI acceptance, another stationary product has lower interface richness and autonomy. This second product is a local minimum of virtual profit after optimizing autonomy.

Fix readiness $x$. Recall that the provider's virtual profit is
\[
w(x;I,\bar q)=\bar q B(x,I)-K(I)-C(\bar q),
\]
where $B(x,I)=D(x,I)-m(x)\mu(x,I)$ and $m(x)=[1-G(x)]/g(x)$. An interior stationary product satisfies the first-order conditions
\[
C'(\bar q)=B(x,I),
\qquad
K'(I)=\bar q B_I(x,I).
\]
Let $(I^\ast,\bar q^\ast)$ be a maximizing product with partial acceptance, $0<\bar q^\ast<1$, and a negative-definite Hessian of $w$ with respect to $(I,\bar q)$.

First, optimize autonomy at each interface:
\[
P(I)=\max_{0\le\bar q\le1}
\{\bar q B(x,I)-C(\bar q)\}-K(I).
\]
Write $\bar q(I)$ for the unique maximizing autonomy at that interface. Strict convexity of $C$ makes $\bar q(I)$ continuous. Whenever it is positive and below one,
\[
C'\bigl(\bar q(I)\bigr)=B(x,I),
\qquad
P'(I)=\bar q(I)B_I(x,I)-K'(I).
\]
Thus a zero of $P'$ in this region satisfies both original first-order conditions.

Second, consider where positive autonomy begins. Under the reliability monotone-hazard condition, $B_I(x,I)>0$ whenever $B(x,I)>0$. For interfaces just above $1/(\kappa x\overline\theta)$, the accepting types have negative virtual contributions, so $B(x,I)<0$. At the maximizing product, however, $B(x,I^\ast)=C'(\bar q^\ast)>C'(0+)$. Continuity therefore gives a smaller interface $I_0<I^\ast$ at which
\[
B(x,I_0)=C'(0+),
\qquad
\lim_{I\downarrow I_0}\bar q(I)=0.
\]
For $I\in(I_0,I^\ast)$, acceptance remains partial and autonomy increases strictly from zero toward $\bar q^\ast$. At the lower boundary,
\[
P'(I_0+)=-K'(I_0)<0.
\]
Virtual profit initially falls because the return to interface investment is multiplied by a delegation share that approaches zero, while the marginal interface cost remains positive.

Third, the shape is different just below the maximizing interface. The first-order conditions give $P'(I^\ast)=0$, and differentiating the optimal autonomy condition yields
\[
P''(I^\ast)
=\bar q^\ast B_{II}(x,I^\ast)-K''(I^\ast)
+\frac{B_I(x,I^\ast)^2}{C''(\bar q^\ast)}<0.
\]
The inequality follows from the negative-definite Hessian. Hence $P'(I)>0$ immediately below $I^\ast$. Since $P'$ is negative just above $I_0$, a minimizer of $P$ on $[I_0,I^\ast]$ lies at some $I_{\min}\in(I_0,I^\ast)$. At this interface,
\begin{equation*}
0<\bar q(I_{\min})<\bar q^\ast,
\qquad
C'\bigl(\bar q(I_{\min})\bigr)=B(x,I_{\min}),
\qquad
K'(I_{\min})=\bar q(I_{\min})B_I(x,I_{\min}).
\end{equation*}
This lower-interface product therefore satisfies both original first-order conditions. However, its virtual profit obeys
\[
P(I_{\min})<P(I_0)=-K(I_0)<0,
\]
so the zero baseline is preferable. It also fails the negative-definite Hessian condition for a regular maximum.

Thus other stationary products can coexist with the unique maximizer under Assumption~\ref{ass:continuity} of the main paper. To establish global optimality, the provider must compare virtual profit across feasible products and with the zero baseline. The comparative statics along a regular interval concern the selected maximizing products.

\end{document}